\documentclass[]{jfm}
\usepackage{multirow}
\usepackage[table]{xcolor}
\usepackage{rotating}
\usepackage{amsmath}
\usepackage{bm}
\usepackage{array}
\usepackage{graphicx}
\usepackage{siunitx}
\usepackage{dcolumn}   
\usepackage{amssymb}
\usepackage{overpic,pict2e}
\usepackage[english]{babel}
\usepackage{natbib}
\usepackage{hyperref}
\usepackage{cleveref}

\crefname{gather}{\cref@equation@name}{\cref@equation@name@plural}
\crefname{eqnarray}{\cref@equation@name}{\cref@equation@name@plural}

\renewcommand{\vec}[1]{\bm{#1}}
\newcommand{\tensor}[1]{\mathsfbi{#1}}

\newcommand{\Zh}{\vec{\hat{z}}}
\newcommand{\Th}{\vec{\hat{t}}}
\newcommand{\Nh}{\vec{\hat{n}}}
\newcommand{\Bh}{\vec{\hat{b}}}
\newcommand{\erho}{\vec{\hat{e}}_{\rho}}
\newcommand{\Rh}{\vec{\hat{R}}}
\newcommand{\R}{\vec{R}}
\newcommand{\norm}[1]{\big|#1\big|}

\shorttitle{Viscous tubular-body theory for plane interfaces}
\shortauthor{L. Koens and B. J. Walker}
\title{Viscous tubular-body theory for plane interfaces}

\author{L. Koens\aff{1}\corresp{\email{l.m.koens@hull.ac.uk}} \and B. J. Walker\aff{2,3}}

\affiliation{\aff{1}Department of Mathematics, University of Hull, Hull HU6 7RX, United Kingdom
\aff{2}Department of Mathematical Sciences, University of Bath, Bath BA2 7AY, United Kingdom
\aff{3}Department of Mathematics, University College London, London, WC1H 0AY, United Kingdom}

\begin{document}

\maketitle

\begin{abstract}
Filaments are ubiquitous within the microscopic world. They occur frequently in both biological and industrial environments and display varied and rich dynamics. Their wide range of applications has spurred the development of a special branch of asymptotics focused on the behaviour of filaments, called slender-body theory. Slender-body theories are typically computationally efficient and focus on the mechanics of an isolated fibre that is not too curved. However, slender-body theories that work beyond these standard limits are needed to explore more complex systems. Recently, we developed tubular-body theory for slow viscous flows, an approach similar to slender-body theory that allows the hydrodynamic traction on any isolated cable-like body in a highly viscous fluid to be determined exactly. In this paper, we extend tubular-body theory to model filaments near plane interfaces by performing an similar expansion on the single-layer boundary integral equations for bodies by a plane interface. In the derivation of the new theory, called tubular-body theory for interfaces, we established a criteria for the convergence of the tubular-body theory series representation, before comparing the result to boundary integral simulations for a prolate spheroid by a wall. The tubular-body theory for interfaces simulations are found to capture the lubrication effects when close to the plane wall. Finally we simulate the hydrodynamics of a helix beneath a free interface and a plane wall to demonstrate the broad applicability of the technique.
\end{abstract}

\section{Introduction}

Fibres and filaments play crucial roles in the motion and organisation of microscopic systems. Many bacteria rotate rigid helical filaments, called flagella, to generate motion \citep{Lauga2016},  some organisms use  microscopic filaments, called cillia, to generate symmetry breaking flows in early embryo development \citep{Hernandez-Pereira2019}, and actin filaments and microtubules play an active role in the organisation of eukaryotic cells \citep{Ganguly2012, Nazockdast2017}. In attempts to mimic their biological conterparts, many microscopic robots also use filaments to control behaviour \citep{Qiu2015, Magdanz2020, Li2021b}, which may lead to the development of new keyhole surgery techniques and methods for targeted drug delivery. The large range of applications of wiry bodies is only possible because of the wide variety of behaviours that a single elastic filament can display \citep{DuRoure2019}.

The sizes and speeds typical of these microscopic cables mean that their movement is dominated by the frictional forces in the surrounding fluid. These filaments can therefore be accurately modelled using the equations for slow viscous flows: the Stokes equations \citep{Kim2005}. However, many numerical approaches struggle to resolve the behaviour of filaments because of their large aspect ratio (defined as length over thickness). This prompted the creation of slender-body theory (SBT), an asymptotic method developed to describe the hydrodynamics of fibres with large aspect ratios. SBTs can be separated into local drag theories \citep{GRAY1955, Koens2021a, Cox} and non-local integral operator theories \citep{Keller1976a, Johnson1979, 1976, Koens2018}.  Local drag theories, sometimes called resistive-force theories (RFTs), provide a linear relationship between the velocity and the force on a filament but require the logarithm of the aspect ratio of the filament to be much larger than one. Resistive-force theories are, therefore, easy to use but only qualitatively describe the behaviour of real filaments.  The non-local, one-dimensional integral operator theories, however, offer greater accuracy \citep{Mori2019, Mori2020, Ohm2019} but need to be solved numerically. This numerical inversion can be tricky, with the most common SBT integral operator being divergent and prone to high-frequency instabilities \citep{Andersson2020}.

Slender-body theory is a powerful tool that has been key in understanding the behaviour of many microscopic systems \citep{Lauga2016, Hernandez-Pereira2019, Ganguly2012, Nazockdast2017, Qiu2015, Magdanz2020, Li2021b, DuRoure2019}. However, most derivations of slender-body theory assume that the fibre is isolated from any other body and that the filament thickness is much smaller than any other length scale within the system. Attempts to overcome these limitations are often very complex \citep{Katsamba2020}, limited to specific regions \citep{Barta1988, Barta1988a, DeMestre1975,  Katz1975}, or to specific geometries \citep{Brennen1977a}. Indeed, slender-body approaches that go beyond these limits  have been identified as a key priority for many interdisciplinary fields \citep{Reis2018, DuRoure2019, Kugler2020}.

The last few years have seen significant developments made in extending SBT beyond the typical limits. Local drag theories have been extended to model fibres in viscoplastic fluids \citep{Hewitt2018} and a RFT model for rods at any distance above a plane interface was found \citep{Koens2021a}. The careful treatment of point torques \citep{Walker2022} and regularised point torques \citep{Maxian2022c} have identified important higher order contributions from rotation. These studies offered new analytical insights into the torques and coupling generated from rotations around a filament's centreline.

Among these developments, we created tubular-body theory (TBT) \citep{Koens2022}. TBT determines the traction jump on any isolated cable-like body, which can be found exactly by iteratively solving a one-dimensional SBT-like operator. Unlike the popular SBT operator of \citet{Johnson1979}, the TBT kernel is compact, symmetric, and self-adjoint, thereby formally transforming the problem into a one-dimensional Fredholm integral equation of the second kind. Fredholm integral equations of the second kind are well posed and there are many techniques to solve them exactly and numerically \citep{Polianin2008}. Though currently a purely numerical tool, TBT is valid well beyond the typical SBT limits, including capturing the hydrodynamics of bodies with arbitrary aspect ratios, thickness variation, and body curvatures.

This paper extends TBT to consider the motion of a cable-like body next to a plane interface. The  geometry of the system is described in \cref{sec:geo} and some background into slow-viscous flows is provided in \cref{sec:stokes}. In \cref{sec:derivation}, the single-layer boundary integral representation for a tubular body by an interface is expanded using the steps of regularisation, binomial series, and reorganisation, similarly to the free-space TBT derivation. Inherited from free-space TBT, the resultant \emph{tubular-body theory by interfaces} (TBTi) system allows for the traction jump on the body to be determined exactly by iteratively solving a well-behaved Fredholm integral equation of the second kind. This iterative representation is equivalent to a geometric series and converges absolutely if certain conditions on the eigenvalues of the operator are met. Using the Galerkin method described in \cref{sec: numerical implementation}, the TBTi equations are solved numerically in \cref{sec:comparison} and compared to both standard and regularised boundary integral simulations for a spheroid with symmetry axis perpendicular to the wall normal. These comparisons highlight the accuracy of TBTi to within numerical tolerance for all the distances and aspect ratios tested, and empirically evidence the satisfaction of the conditions placed on the TBTi operator. In particular, these examples highlight how TBTi is able to accurately capture lubrication effects, though additional iterations are required as an object closely approaches a boundary. Finally, in \cref{sec:helix}, we compare the traction jump associated with a helix approaching a rigid wall to that near a free interface, each of which are found to be consistent with the scaling of lubrication forces.

\section{Geometry of the tubular body} \label{sec:geo}

\begin{figure}
    \centering
   \includegraphics[width=0.9\textwidth]{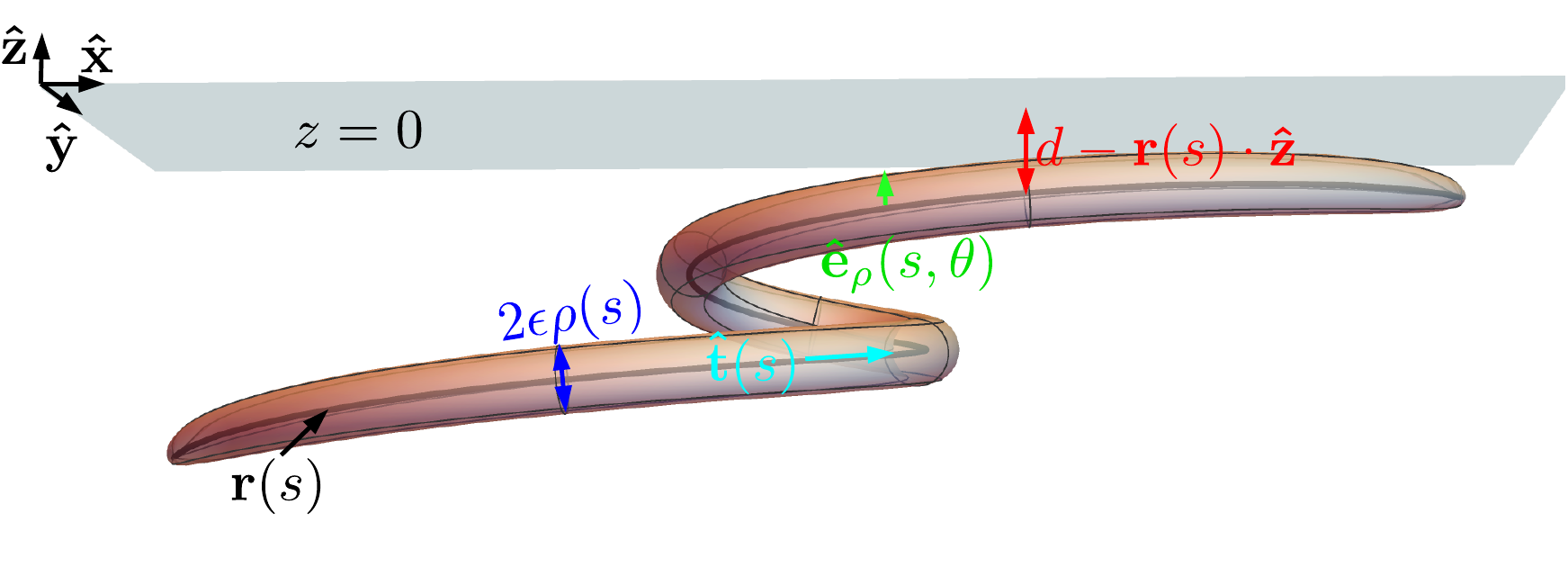}
    \caption{Diagram of a tubular body under a plane interface at $z=0$. The distance from the place interface is denoted by $d$, $\vec{r}(s)$ represents the centreline of the tubular body, $\epsilon \rho(s)$ is the thickness of the body at $s$, $\vec{\hat{t}}(s)$ is the tangent vector to the centreline, and $\erho(s,\theta)$ is the local radial vector around the centreline.}
    \label{fig:diagram}
\end{figure}

The surface of a tubular body is geometrically identical to that of a slender body but does not assume that the aspect ratio of the body is large. Beneath a plane interface, such a body can be parameterised by an arclength parameter $s\in[-1,1]$ and an angular parameter $\theta$ as
\begin{equation}
\vec{S}(s,\theta) = \vec{r}(s) + \epsilon \rho(s) \erho - d \Zh\,,
\end{equation}
where $\vec{r}(s)$ is the centreline of the filament, $\epsilon$ is the maximum thickness of the filament (inverse aspect ratio in unscaled units), $\epsilon \rho(s)$ is the cross-sectional radius ($0\leq\rho(s)\leq1$), $\erho = \cos[\theta-\theta_{i}(s)]\Nh(s) +\sin[\theta-\theta_{i}(s)]\Bh(s)$, and $d$ is the offset of the body from the plane interface located at $z=0$ (\cref{fig:diagram}). In the above, $\Zh$ is the unit vector in the direction of increasing $z$, $\Nh(s) $ is the normal vector of the centreline, $\Bh(s)$ is the binormal vector of the centreline, and $\theta_{i}(s)$ sets the origin of the $\theta$ coordinate. The function $\theta_{i}$ is defined such that $\mathrm{d}\theta_i/\mathrm{d}s = \tau(s)$ for torsion $\tau=\mathrm{d}\Bh/\mathrm{d}s\cdot\Nh$, which removes any dependence of our analysis on the torsion \citep{Koens2018}. We assume that the tubular body lies completely under the $z=0$ plane and it does not intersect itself, so that $\vec{S}(s,\theta)\cdot\Zh<0$ and $\vec{S}(s,\theta)\neq\vec{S}(s',\theta')$ if $(s,\theta)\neq(s',\theta')$, respectively.

\section{Stokes flow and the Green's function for a plane interface} \label{sec:stokes}

The slow viscous flow around a tubular body can be accurately modelled by the incompressible Stokes equations \citep{Kim2005}
\begin{gather}
    \mu \nabla^2 \vec{u} - \nabla p = \vec{0}\,, \label{S1}\\
     \nabla \cdot \vec{u} = 0\,, \label{S2}
\end{gather}
where $\mu$ is the dynamic viscosity of the fluid, $\vec{u}$ is the fluid velocity, and $p$ is the fluid pressure. The drag force, $\vec{F}$, and torque, $\vec{L}$, on the fluid from the tubular body are 
\begin{gather}
   \vec{F} = \iint_{S} (\bm{\sigma} \cdot \vec{\hat{n}}) \,dS\,,  \label{force}\\
   \vec{L} = \iint_{S} \vec{S} \times (\bm{\sigma} \cdot \vec{\hat{n}})  \,dS\,, \label{torque}
\end{gather}
where the integrals are taken over the surface of the body, $\vec{\hat{n}}$ is the outward pointing unit normal to the surface, and $\bm{\sigma}  = - p \tensor{I} + \mu ( \nabla \vec{u} + (\nabla \vec{u})^T)$ is the fluid stress tensor.

The incompressible Stokes equations are linear and time independent, with the flow therefore depending only on the instantaneous geometry of the system and any boundary conditions. Hence, the drag force and torque on the fluid from rigid body motion can always be written as
\begin{equation}
    \begin{pmatrix}
        \vec{F}  \\
         \vec{L}
    \end{pmatrix} = \begin{pmatrix}
        \tensor{R}^{FU} & \tensor{R}^{F\Omega} \\
         (\tensor{R}^{F\Omega})^{T} & \tensor{R}^{L\Omega}
    \end{pmatrix} \begin{pmatrix}
        \vec{U}  \\
         \bm{\Omega}
    \end{pmatrix},
\end{equation}
where $\vec{U}$ is the linear velocity of the body, $\bm{\Omega}$ is the angular velocity and $\tensor{R}^{FU}$, $\tensor{R}^{F\Omega}$ and $\tensor{R}^{L\Omega}$ are $3\times 3$ matrices describing the drag force generated from translation, the drag force generated from rotation (or, equivalently, the torque generated from translation), and the torque generated from rotation, respectively. 

 Exact solutions of the incompressible Stokes equations, \cref{S1,S2}, only exist for simple geometries \citep{Kim2005}. As a result, most solutions are found asymptotically or numerically. Many of these asymptotic and numerical methods rely on the Green's function solution for the Stokes equations, called the Stokeslet. The Stokeslet represents the flow from a point force of strength $\vec{f}$ on the fluid that is located at $\vec{y}$. The flow from a Stokeslet, $\vec{u}_S$, satisfies
\begin{equation}
    \mu \nabla^2 \vec{u}_S - \nabla p = \vec{f} \delta(\vec{x}-\vec{y}).
\end{equation}
along with the incompressibility condition of \cref{S2}.
In free space, it is given explicitly as
\begin{equation}
8 \pi \mu \vec{u}_S(\vec{x}) =\tensor{G}_{S}(\vec{R}) \cdot \vec{f}\,, \quad \tensor{G}_{S}(\vec{R}) = \frac{\tensor{I}+ \Rh\Rh}{|\R|}\,,
\end{equation}
where $\vec{x}$ is a point in the domain and we define $\vec{R}= \vec{x}-\vec{y}$ as the vector from the point force to the point of interest in the flow \citep{Kim2005}. Here and throughout, $\Rh = \vec{R} / \norm{\vec{R}}$, $\hat{\cdot}$ denotes a unit-normalised vector, and $\norm{\cdot}$ denotes the length of the vector.

The Stokeslet for more complicated geometries can be constructed using the representation by fundamental singularities. This method places Stokeslets and their derivatives outside the fluid region such that the boundary conditions are satisfied. Such a representation is always theoretically possible for the flow around any body \citep{Kim2005}, but the location and strengths of the singularities are often not known a priori. 
However, the flow due to a point force under a plane interface at $z=0$ is known. In particular, if the fluid beneath the interface has viscosity $\mu_1$ and the fluid above the interface has viscosity $\mu_2$, the solution can be found by placing adding a Stokeslet, a force dipole (the derivative of the Stokeslet with respect to its position), and source dipole (Laplacian of the Stokeslet) in the fluid region above the interface \citep{Aderogba1978}. The resultant flow $\vec{u}_{S}^{*}$ in the lower fluid region is therefore given by
\begin{equation}
8 \pi \mu_{1} \vec{u}_S^{*}(\vec{x}) = \tensor{G}_{S}(\vec{R}) \cdot \vec{f} + \tensor{G}_{S}^{*}(\R') \cdot \vec{f}\,,
\end{equation}
where $\lambda= \mu_2/\mu_1$ is the viscosity ratio of the two fluids, $y_z=\vec{y} \cdot \Zh < 0$,
\begin{align}
\tensor{G}_{S}^{*}(\R') &=    \frac{\tensor{I}+ \Rh' \Rh'}{|\R'|}\cdot \tensor{B} -  \frac{2 \lambda}{1 + \lambda} y_z \left[(\R'\cdot\Zh-y_z)\frac{\tensor{I} - 3 \Rh'\Rh'}{|\R'|^3} + \frac{\R' \Zh - \Zh \R'}{|\R'|^3} \right]\cdot \tensor{A}\,, \\
 \tensor{B} &= \frac{1 - \lambda}{1 + \lambda}\left(\tensor{I}- \Zh\Zh\right) -\Zh\Zh\,, \\
 \tensor{A} &= \tensor{I}- 2\Zh\Zh\,, \\
\R' &= \vec{x} - \tensor{A} \cdot \vec{y}\,.
\end{align}
In the above, $\tensor{A}$ is the reflection matrix across the $z=0$ plane. This solution for the flow due to a Stokeslet underneath a plane interface represents the flow under a free surface when $\lambda=0$ and a rigid wall in the limit $\lambda \to \infty$.

The Stokeslet plays an important role developing numerical and asymptotic solutions to the incompressible Stokes equations, \cref{S1,S2}. Several asymptotic theories use a representation-by-fundamental-singularities approach to construct approximate solutions for the flow around bodies with special symmetries \citep{Keller1976a, Johnson1979}. For example, some slender-body theories (SBTs) approximate the flow around an isolated slender filament by placing Stokeslets and source dipoles placed along the centreline of the fibre \citep{Johnson1979}. The strength of the Stokeslets and source dipoles are determined by asymptotically expanding the no-slip boundary condition in the inverse aspect ratio of the body. This expansion sets a linear relationship between the strength of the Stokeslets and the source dipoles and relates the Stokeslet strength to the centreline velocity, $\vec{U}_{c}(s)$, through a one-dimensional integral equation given by
\begin{multline}
    8 \pi \mu \vec{U}_{c}(s) =\int\limits_{-1}^{1}   \left( \frac{\tensor{I}+ \vec{\hat{R}}_{0}\vec{\hat{R}}_{0}}{|\vec{R}_{0}|} \cdot \vec{q}(s') -\frac{\tensor{I}+ \vec{\hat{t}}\vec{\hat{t}}}{|s'-s|} \cdot  \vec{q}(s) \right)\,d s'  \\{}+ \left[L_{SBT} (\tensor{I}+ \vec{\hat{t}}\vec{\hat{t}})+ \tensor{I}- 3 \vec{\hat{t}}\vec{\hat{t}} \right]\cdot\vec{q}(s)\,,
\end{multline}
where $\vec{R}_0(s,s') = \vec{r}(s)-\vec{r}(s')$ is a vector between two points on the centreline of the bod, $\vec{\hat{t}}(s) = \partial_s \vec{r}(s)$ is the tangent to the centreline, $\vec{q}(s)$ is the Stokeslet strength, and $L_{SBT} =\ln[4 (1-s^{2})/(\epsilon^{2}\rho^{2}(s))]$. Though structurally similar to a one-dimensional Fredholm integral equation of the second kind, this equation does not share the same properties due to the kernel being singular, thereby making it difficult to solve \citep{Tornberg2004, Tornberg2020}. Even so, this formulation has been used successfully in varied circumstances \citep{Lauga2016, Hernandez-Pereira2019, Ganguly2012, Nazockdast2017, Qiu2015, Magdanz2020, Li2021b, DuRoure2019} and derived in many different ways \citep{Keller1976a,Koens2018}. Extensions of SBT to include boundaries tend to only apply in limited regimes \citep{Barta1988a, Lisicki2016b, Brenner1962,JEFFREY1981, DeMestre1975} or for a limited set of geometries \citep{Koens2021a, Man2016}.

Most numerical approaches to solve the incompressible Stokes equations use the Green's function nature of the Stokeslet to transform the equations into the boundary integrals \citep{Kim2005, Pozrikidis1992}
\begin{multline}\label{eq: boundary integral equations}
4 \pi \mu \vec{U}(\vec{x}) =  \iint_S \,d S(\vec{x}_{0}) \left[ \tensor{G}(\vec{x}-\vec{x}_0) \cdot\vec{f}(\vec{x}_{0})\right]  \\{}+ \mu \iint_S^{PV} \,d S(\vec{x}_{0}) \left[\vec{U}(\vec{x}_{0})\cdot \tensor{T}(\vec{x}-\vec{x}_0) \cdot \vec{\hat{n}}(\vec{x}_{0}) \right]\,, 
\end{multline}
where all the integrals are carried out over the boundaries of the system, $\vec{U}(\vec{x})$ is the velocity on the surface, $\vec{\hat{n}}(\vec{x}_{0})$ is the surface normal pointing into the fluid, $\vec{f}(\vec{x}_{0}) = \bm{\sigma}(\vec{x}_{0}) \cdot \vec{\hat{n}}(\vec{x}_{0})$ is the surface traction, $\tensor{T}(\vec{R})$ is the stress generated from the Stokeslet, and the superscript $PV$ denotes a principal value integral. The boundary integrals are exact and apply for any geometry in which the Green's function, $\tensor{G}$, is known \citep{Pozrikidis1992}. If the volume of the tubular-body is constant, this equation can be transformed into the single-layer boundary integral 
\begin{equation}
8 \pi \mu \vec{U}(\vec{x}) =  \iint_S \,d S(\vec{x}_{0}) \tensor{G}(\vec{x}-\vec{x}_0) \cdot\vec{\tilde{f}}(\vec{x}_{0})\,, 
\end{equation}
where $\vec{\tilde{f}}(\vec{x}_{0})$ represents the jump in surface traction. Notably, the force and torque over any closed surface can be found identically to \cref{force,torque} but with $\vec{\tilde{f}}(\vec{x}_{0})$ replacing the traction \citep{Pozrikidis1992}. These equations have been studied extensively and can be solved by discretizing the surface and assuming a known behaviour of the traction over this region \citep{Pozrikidis1992}. However, this approach often struggles when modelling slender bodies because of the large aspect ratios of these systems.

Since the single-layer boundary integral represents the flow exactly in these circumstances \citep{Kim2005}, we can use it to develop a tubular-body theory for interfaces (TBTi). Unlike other expansions of the boundary integrals \citep{Koens2018}, the TBT approach creates a similar one-dimensional slender-body theory integral operator, but with a compact, symmetric, and self-adjoint kernel. Furthermore the iterative solving of this operator can be used to reconstruct the jump in surface traction exactly. This overcomes several of the numerical issues encountered in SBT and allows TBTi to surpass the limitations of typical SBTs.

 \section{Tubular-body theory for interfaces} \label{sec:derivation}
 
 Tubular-body theory seeks to transform the single-layer boundary integral representation into a series of well-behaved one-dimensional Fredholm integral equations of the second kind that can be sequentially inverted to determine higher order corrections. The structure of these integral equations is similar to the classical SBT formalism, but avoids several of their issues and can handle arbitrary aspect ratios. Though the focus of this work is on tubular bodies by plane interfaces, the development of this approach is easily generalised to other scenarios where Green's functions are available. We have presented our formulation in a manner that highlights this.
 
 \subsection{Regularisation of the boundary integrals.}
 
The single-layer boundary integral representation for a tubular body by an interface can always be expressed as 
\begin{equation}
 8 \pi \mu_1 \vec{U}(\vec{S}(s,\theta)) =  \int\limits_{-1}^{1} \,d s' \int\limits_{-\pi}^{\pi} \,d \theta' \tensor{G}(s,\theta,s',\theta') \cdot \vec{\bar{f}}(s',\theta') , \label{BI}
\end{equation}
where $\vec{U}(\vec{S}(s,\theta))$ is the known velocity at $\vec{S}(s,\theta)$ on the surface of the body, $\tensor{G}(s,\theta,s',\theta') =\tensor{G}_{S}(\vec{S}(s,\theta)-\vec{S}(s',\theta')) + \tensor{G}_{S}^{*}(\vec{S}(s,\theta)-\tensor{A}\cdot\vec{S}(s',\theta'))$ is the Green's function for the flow at $\vec{S}(s,\theta)$ from a point force located at $\vec{S}(s',\theta')$, and $\vec{\bar{f}}(s',\theta')$ is the unknown surface traction jump, $\vec{\tilde{f}}$, multiplied by the corresponding surface element at $(s',\theta')$.
The integrand of the boundary integrals diverges as $(s',\theta') \to (s, \theta)$ because the free space component of the Green's function, $\tensor{G}_{S}(\vec{S}(s,\theta)-\vec{S}(s',\theta'))$, blows up at this location. The interface corrections $\tensor{G}_{S}^{*}(\vec{S}(s,\theta)-\tensor{A}\cdot\vec{S}(s',\theta'))$ are non-singular if $d>0$. The divergence of the free space Green's function is formally not an issue, though it can cause problems when asymptotically expanding the integrand.

There are numerous ways to regularise boundary integral representations to overcome the singularity of the free-space kernel \citep{Cortez2005, Klaseboer2012, Batchelor2006}. One of the simplest is by adding and subtracting an existing solution to the boundary integral representation chosen such that the integrands cancel when $(s',\theta') \to (s, \theta)$. For example, $\vec{\bar{f}}$ is known to be constant for the rigid body translation of an isolated spheroid \citep{Brenner1963, Martin2019} parameterised as
\begin{equation}
\vec{S}_{e}(s,\theta) = a s \vec{\hat{x}}' + \epsilon c \sqrt{1-s^2} \bm{\hat{\rho}}(\theta) +\vec{q}, 
\end{equation}
where $a$ and $\epsilon c$ are the semi axes of the spheroid, $\vec{\hat{x}}'$ is a unit vector along the symmetry axis, $\bm{\hat{\rho}}(\theta)$ is the radial director perpendicular to the symmetry axis, and $\vec{q}$ is the centre of the spheroid. The solution of \citet{Brenner1963} gives
\begin{equation}
 \tensor{M}_{A}\cdot\vec{\bar{f}}(s,\theta)   =  \int\limits_{-1}^{1} \,d s' \int\limits_{-\pi}^{\pi} \,d \theta' \tensor{G}_{S}(\vec{S}_{e}(s,\theta)-\vec{S}_{e}(s',\theta')) \cdot \vec{\bar{f}}(s,\theta) , \label{BIspheroid}
\end{equation}
where $\alpha = \epsilon c/a$ is the inverse aspect ratio of the spheroid. The matrix $\tensor{M}_A$ is proportional to the translational mobility matrix of the spheroid and equals
\begin{align}
 \tensor{M}_{A} &= \zeta_{\parallel} \vec{\hat{x}}' \vec{\hat{x}}' + \zeta_{\perp} (\tensor{I} -\vec{\hat{x}}' \vec{\hat{x}}')\,, \\
 \frac{a \beta^{3/2}}{4 \pi \mu_1} \zeta_{\parallel} &= (\beta-1) \arccos(\alpha^{-1}) + \sqrt{\beta}\,, \\
  \frac{a \beta^{3/2}}{2 \pi \mu_1} \zeta_{\perp} &= (3\beta+1) \arccos(\alpha^{-1}) - \sqrt{\beta}\,, \\
  \beta &= \alpha^2-1\,.
\end{align}  
 This boundary integral result for a spheroid can be used to regularise the free space Green's function in the boundary integrals for the unknown body $\vec{S}$, if the surface of the spheroid $\vec{S}_e$ is chosen to be similar to that of the tubular body as $(s',\theta') \to (s, \theta)$. Since the spheroid parametrisation has four independent parameters ($a$, $\vec{\hat{x}}'$, $\epsilon c$, $\vec{q}$), it is possible to enforce that the position and the tangent plane to the tubular body at $(s,\theta)$ agree with those of the spheroid at $(s_e,\theta)$, where $s_e$ is the location along the arc length of the spheroid at which the surface and tangent plane matches. Explicitly, we require $\vec{S}_e(s_e,\theta) = \vec{S}(s,\theta)$, $\partial_{s_e}\vec{S}_e(s_e,\theta) =\partial_{s} \vec{S}(s,\theta)$ and $\partial_{\theta}\vec{S}_e(s_e,\theta) =\partial_{\theta} \vec{S}(s,\theta)$. These conditions give the following relationships between the surface of the body and the parametrisation of the spheroid:
 \begin{align}
 \vec{\hat{x}}' &= \vec{\hat{t}}(s)\,, \label{r1}\\
 \bm{\hat{\rho}}(\theta) &= \erho(s,\theta)\,, \label{r2}\\
 \vec{q} +a s_e \vec{\hat{x}}'&= \vec{r}(s)\,, \label{r3}\\
 2 c^2 &= \rho^2(s) + \rho(s) \sqrt{\rho^2(s)+4(\partial_s \rho(s))^2}\,, \label{r4}\\
 a &= 1 - \vec{\hat{t}}(s) \cdot \partial_s \erho(s,\theta)\,, \label{r5}\\
 c^2 s_e &= \rho(s) \partial_s \rho(s)\,, \label{r6}
 \end{align}
where $ \vec{\hat{t}}(s) = \partial_{s} \vec{r}(s)$ is the tangent to the centreline of the body.

With these choices, adding and subtracting \cref{BIspheroid} from \cref{BI} regularises the boundary integral for our tubular body, leading to
\begin{align}
 8 \pi\mu_1 \vec{U}(\vec{S}(s,\theta)) &=    \int\limits_{-1}^{1} \,d s' \int\limits_{-\pi}^{\pi} \,d \theta' \tensor{G}_{S}(\vec{S}(s,\theta)-\vec{S}(s',\theta')) \cdot \vec{\bar{f}}(s',\theta') \notag\\
  &+\int\limits_{-1}^{1} \,d s' \int\limits_{-\pi}^{\pi} \,d \theta' \tensor{G}_{S}^{*}(\vec{S}(s,\theta)-\tensor{A}\cdot\vec{S}(s',\theta')) \cdot \vec{\bar{f}}(s',\theta')\notag\\
  &-  \int\limits_{-1}^{1} \,d s' \int\limits_{-\pi}^{\pi} \,d \theta' \tensor{G}_{S}(\vec{S}_{e}(s_e,\theta)-\vec{S}_{e}(s',\theta')) \cdot \vec{\bar{f}}(s,\theta) \notag\\
   &+ \tensor{M}_{A}\cdot\vec{\bar{f}}(s,\theta) \,.\label{regG} 
\end{align} 
We note that $\vec{S}_e$, $s_e$, and $\tensor{M}_{A}$ depend on $s$ and $\theta$ via \crefrange{r1}{r6}. The matching of the two surfaces means that the singularity in the first integrand as $(s',\theta') \to (s,\theta)$ now precisely cancels with the singularity in the third integrand as  $(s',\theta') \to (s_e,\theta)$.

\subsection{Identifying the leading-order terms}
 
 The next step is to manipulate the regularised boundary integrals in \cref{regG} to find the leading-order terms. These leading-order terms will act as the SBT-like operator in the tubular-body theory expansion. In keeping with the SBT approach, the leading terms should be structurally equivalent to a Fredholm integral equation of the second kind, as these are well-posed problems and have been studied extensively \citep{Polianin2008}. This requires the expansion process to somehow allow the evaluation of the $\theta'$ integration within \cref{regG} while keeping the expanded Green's function (the kernel) compact. Additionally, it will be useful if the kernel is symmetric and self-adjoint, as the operator will have real eigenvalues and additional desirable properties \citep{Polianin2008}.
 
Notably, the integration over $\theta'$ can be evaluated if all the $\theta'$ terms within the denominator of the Green's function are moved to the numerator in the expansion process \citep{Koens2018}. If done through a Taylor series of expansion in the inverse aspect ratio $\epsilon$, which here we don't assume is small, this recovers the classical slender-body theory equations. The kernel of these equations is, however, not-compact. Recently there have been many attempts have been made to fix this \citep{Shi2022, Tatulea-Codrean2021, Andersson2020, Walker2022, Walker, Maxian2022b}.

In contrast to SBT, the tubular-body theory derivation creates a compact, symmetric, and self-adjoint kernel by expanding each denominator in the Green's function using the binomial series. This expansion converges absolutely whenever $(s,\theta) \neq (s',\theta')$, irrespective of the body geometry or position. In the previous TBT derivation, this was done using a single binomial expansion, motivated by an erroneous claim about the triangle inequality. Here, we correct this by applying the binomial series twice. The final structure, however, remains the same.

First, one writes the argument of the free space Green's function as
\begin{equation}
\vec{S}(s,\theta)-\vec{S}(s',\theta') = \vec{R}_0(s,s') + \epsilon \Delta  \erho(s,\theta,s',\theta') \,, \label{3}
\end{equation}
where $\vec{R}_0(s,s') = \vec{r}(s)-\vec{r}(s')$ is a vector between two points on the centreline of the body and $\Delta  \erho(s,\theta,s',\theta') = \rho(s) \erho(s,\theta) -\rho(s') \erho(s',\theta')$ is the difference between the cross-section vectors at $(s,\theta)$ and $(s',\theta')$. The length squared of the argument is therefore
\begin{equation}
    \norm{\vec{S}(s,\theta)-\vec{S}(s',\theta')}^2 = \norm{\vec{R}_0(s,s')}^2 + \epsilon^2 \norm{\Delta  \erho(s,\theta,s',\theta')}^2 \\
+ 2\epsilon  \vec{R}_0(s,s') \cdot \Delta  \erho(s,\theta,s',\theta')\,,
\end{equation}
where the first two terms are the squared lengths of each of the vectors in \cref{3} while the last term is the cross term. This equation can be rewritten as
\begin{equation}
    \norm{\vec{S}(s,\theta)-\vec{S}(s',\theta')}^2 = \left[\norm{\vec{R}_0(s,s')}^2 + \epsilon^2 \norm{\Delta  \erho(s,\theta,s',\theta')}^2 \right]\left[1+ R_{\Delta }^{(1)}(s,\theta,s',\theta')\right]\,, 
\end{equation}
where
\begin{equation}\label{eq: def of delta R1}
   \left[\norm{\vec{R}_0(s,s')}^2 + \epsilon^2 \norm{\Delta  \erho(s,\theta,s',\theta')}^2 \right]  R_{\Delta }^{(1)}(s,\theta,s',\theta') = 2\epsilon  \vec{R}_0(s,s') \cdot \Delta  \erho(s,\theta,s',\theta')\,.
\end{equation}
The size of $R_{\Delta }^{(1)}(s,\theta,s',\theta')$ can be bound with the triangle inequality. The triangle inequality implies that for any two vectors $\vec{a}$ and $\vec{b}$, $\norm{\vec{a}}^2 + \norm{\vec{b}}^2 \geq 2 \norm{\vec{a} \cdot \vec{b}}$, with equality holding if and only if $\vec{a}=\pm\vec{b}$.  If $\vec{a} = \vec{R}_0(s,s')$ and $\vec{b} = \epsilon\Delta  \erho(s,\theta,s',\theta')$, $\vec{a}=\pm\vec{b}$ can only occur if the tubular-body intersects itself.
Hence, provided that the body does not self intersect, the triangle inequality shows that $\norm{R_{\Delta}^{(1)}(s,\theta,s',\theta')} < 1$ for all $(s',\theta')$. This same bound does not hold when considering the cross terms that result from a sum of three vectors, rather than two. This was the erroneous assumption that led to the error in the original TBT derivation, though this does not impact on the derived formalism.

The bound on $R_{\Delta}^{(1)}(s,\theta,s',\theta')$ prompts the denominator of each of the terms within the free-space Green's function to be written as
\begin{multline}
\norm{\vec{S}(s,\theta)-\vec{S}(s',\theta')}^{-2n} =  \\ \left[\norm{\vec{R}_0(s,s')}^2 + \epsilon^2 \norm{\Delta  \erho(s,\theta,s',\theta')}^2\right]^{-n} \left[1+ R_{\Delta}^{(1)}(s,\theta,s',\theta')\right]^{-n}, 
\end{multline}
which is structurally equivalent to a binomial series. Generally, a binomial series can be written as 
\begin{equation}
(1+x)^\alpha = \sum_{k=0}^{\infty}\binom{\alpha}{k}x^k,
\end{equation} 
where the generalised binomial coefficient is given by
\begin{equation}
\binom{\alpha}{k}  =  \frac{1}{k!} \prod_{n=0}^{k+1} (\alpha-n)\,. \label{binomial}
\end{equation}
The binomial series converges absolutely if $\norm{x}<1$ and $\alpha \in \mathbb{C}$. Therefore, taking $\alpha=-n$ and $x = R_{\Delta}^{(1)}(s,\theta,s',\theta')$, the denominators in the free-space Green's function can be expressed as
\begin{multline}
    |\vec{S}(s,\theta)-\vec{S}(s',\theta')|^{-2n} =\\  \left[\norm{\vec{R}_0(s,s')}^2 + \epsilon^2 \norm{\Delta  \erho(s,\theta,s',\theta')}^2\right]^{-n} \sum_{k_1=0}^{\infty}\binom{-n}{k_1} R_{\Delta}^{(1)}(s,\theta,s',\theta')^{k_1}\,. \label{step1}
\end{multline}

This first binomial series moves the $\theta'$ that is related to the dot product of $\vec{R}_0(s,s')$ and $\Delta  \erho(s,\theta,s',\theta')$ from the denominator to the numerator of the Green's function. However, $\theta'$ dependence remains within the $\norm{\Delta  \erho(s,\theta,s',\theta')}^2$ term of the denominator. 

The $\theta'$ dependence that remains within the denominator can be addressed with a second binomial series. Similarly to the first expansion, the length squared of  $\Delta  \erho(s,\theta,s',\theta')$ can be written as
\begin{eqnarray}
    \norm{\Delta  \erho(s,\theta,s',\theta')}^2  = \rho^2(s) +\rho^2(s') - 2  \rho(s) \rho(s') \erho(s,\theta) \cdot \erho(s',\theta')\,,
\end{eqnarray}
allowing the remaining terms in the denominator to be expressed as
\begin{equation}
\norm{\vec{R}_0(s,s')}^2 + \epsilon^2 \norm{\Delta  \erho(s,\theta,s',\theta')}^2 = \norm{\tilde{\vec{R}}(s,s')}^2 \left[1+ R_{\Delta}^{(2)}(s,\theta,s',\theta')\right]\,, 
\end{equation}
where
\begin{align}
\norm{\tilde{\vec{R}}(s,s')}^2 &= \norm{\vec{R}_0(s,s')}^2 + \epsilon^2 \rho^2(s) + \epsilon^2 \rho^2(s')\,, \\
\norm{\tilde{\vec{R}}(s,s')}^2 R_{\Delta}^{(2)}(s,\theta,s',\theta') &=  - 2 \epsilon^2 \rho(s) \rho(s') \erho(s,\theta) \cdot \erho(s',\theta')\,.
\end{align} 
Similarly to the first expansion, the triangle inequality tells us that $  \rho^2(s) + \rho^2(s') \geq 2  \rho(s) \rho(s') \norm{   \erho(s,\theta) \cdot \erho(s',\theta')}$. This means that $\norm{R_{\Delta}^{(2)}(s,\theta,s',\theta')}< 1$ for $s \neq s'$ because the distance between any two points on the centerline is greater than zero, $\norm{\vec{R}_0(s,s')} >0$, when $s \neq s'$. If $s=s'$, the distance between points on the centreline goes to zero, so that $\norm{\vec{R}_0(s,s')} =0$ and the triangle inequality becomes $ 1 \geq \norm{   \erho(s,\theta) \cdot \erho(s,\theta')}$. Hence, $\norm{R_{\Delta}^{(2)}(s,\theta,s,\theta')}=1$ if the local radial vector at $(s,\theta)$, $\erho(s,\theta)$, is parallel to the vector at $(s,\theta')$, $\erho(s,\theta')$. These local radial vectors are parallel if $\theta=\theta'+m \pi$, where $m$ is an integer, meaning that $R_{\Delta}^{(2)}(s,\theta,s',\theta')<1$ if $(s,\theta) \neq (s', \theta'+m\pi)$. A binomial series in $R_{\Delta}^{(2)}(s,\theta,s',\theta')$, therefore, allows us to express the denominators as
\begin{multline}
|\vec{S}(s,\theta)-\vec{S}(s',\theta')|^{-2n} =  \\ |\tilde{\vec{R}}(s,s')|^{-2n} \sum_{k_1=0}^{\infty} 
\binom{-n}{k_1} R_{\Delta}^{(1)}(s,\theta,s',\theta')^{k_1} \sum_{k_2=0}^{\infty} \binom{-n}{k_2} R_{\Delta}^{(2)}(s,\theta,s',\theta')^{k_2}, 
\end{multline}
if $(s,\theta) \neq (s',\theta'+m \pi)$. Geometrically, $|\tilde{\vec{R}}(s,s')|^2$ is the total squared lengths of $\vec{R}_0(s,s')$, $\epsilon \rho(s)\vec{\hat{e}}_\rho(s,\theta)$  and $\epsilon \rho(s')\vec{\hat{e}}_\rho(s',\theta')$, while the $R_{\Delta}^{(i)}(s,\theta,s',\theta')$ contain the interactions between the vectors, where $i=1,2$. 

The summation over $k_2$ does not converge when $(s,\theta) = (s',\theta'+m \pi)$. However, these points are treated with the regularisation of the boundary integrals. The spheroid used in the regularised boundary integrals, \cref{regG}, was chosen to mimic the dimensions and tangent plane of the tubular body at $(s,\theta)$. This means that the radius and radial directors $\erho$ of the spheroid match with the body at this location. Hence, when $s=s'$, the terms $\norm{\tilde{\vec{R}}}$, $R_{\Delta}^{(1)}(s,\theta,s,\theta')$,  and $R_{\Delta}^{(2)}(s,\theta,s,\theta')$ are the same form for the tubular body and the regularising spheroid. The subtraction of the spheroid geometry in the regularised boundary integrals, \cref{regG}, causes each term of the binomial expanded free-space kernel for the tubular body to cancel with its counterpart from the regularising spheroid, removing the convergence issue when $(s,\theta) = (s',\theta'+m \pi)$. This is by construction.

The applications of sequential binomial series allows the free-space Green's function to be rewritten as
 \begin{equation}
 \tensor{G}_{S}(\vec{S}(s,\theta)-\vec{S}(s',\theta')) = \tensor{K}_{S}(s,s')+O( R_{\Delta}^{(i)}(s,\theta,s',\theta'))
 \end{equation}
 for $i=1,2$, where $\tensor{K}_{S}(s,s')$ is the leading-order kernel and equals
 \begin{equation}
 \tensor{K}_{S}(s,s') =\frac{\tensor{I}}{\norm{\tilde{\vec{R}}}} + \frac{\vec{R}_0\vec{R}_0}{\norm{\tilde{\vec{R}}}^3}\,.
 \end{equation} 
 The leading-order approximation for the integration of the free space Green's function therefore becomes
 \begin{align}
 \int\limits_{-1}^{1} \,d s' \int\limits_{-\pi}^{\pi} \,d \theta' \tensor{G}_{S}(\vec{S}(s,\theta)-\vec{S}(s',\theta')) \cdot \vec{\bar{f}}(s',\theta')  &\approx \int\limits_{-1}^{1} \,d s' \int\limits_{-\pi}^{\pi} \,d \theta' \tensor{K}_{S}(s,s') \cdot \vec{\bar{f}}(s',\theta') \notag \\
 &= 2 \pi \int\limits_{-1}^{1} \,d s'  \tensor{K}_{S}(s,s') \cdot \langle \vec{\bar{f}}(s',\theta') \rangle_{\theta'}\,, \label{sing_ex}
 \end{align}
 where $\langle \cdot \rangle_{\theta'} = \int\limits_{-\pi}^{\pi} \,d \theta' / (2 \pi)$.  Hence, the binomial expansion has effectively treated the $\theta'$ integration and left a well-behaved integrand. This is the same kernel as found via an erroneous method in the free space TBT formalism \citep{Koens2022}.

 The expansion of the free space Green's function naturally includes the regularising spheroid geometry. The result of the regularising spheroid integral can, therefore, be found by recognising that, for the spheroid, $\vec{r}(s) \equiv a s \vec{\hat{x}}$ and $\rho(s) \equiv c \sqrt{1-s^2}$. Hence, the binomial series give
 \begin{equation}
 \tensor{G}_{S}(\vec{S}_e(s_e,\theta)-\vec{S}_e(s',\theta')) = \tensor{K}_{S,e}(s_e,s')+O( R_{\Delta}^{(i)}(s_e,\theta,s',\theta'))\,,
 \end{equation}
 where 
 \begin{align}
 \tensor{K}_{S,e}(s_e,s') &=\frac{\tensor{I}}{\norm{\tilde{\vec{R}}_e}} + a^2(s,\theta) (s_e(s)-s')^2\frac{\vec{\hat{t}}(s)\vec{\hat{t}}(s)}{\norm{\tilde{\vec{R}}_e}^3} \\
 \norm{\tilde{\vec{R}}_e(s_e,\theta,s')}^2 &= a^2(s,\theta)(s_e(s)-s')^2 + \epsilon^2 c^2(s) (2 - s_e^2(s) - s'^2).
 \end{align}
The above explicitly includes the additional $(s,\theta)$ dependence in $s_e(s)$, $a(s,\theta)$ and $c(s)$ as dictated by \crefrange{r1}{r6}. The leading-order approximation for the integration of the spheroid's Green's function therefore becomes
 \begin{align}\label{eq: leading order integral equation}
 \int\limits_{-1}^{1} \,d s' \int\limits_{-\pi}^{\pi} \,d \theta' \tensor{G}_{S}(\vec{S}_e(s_e,\theta)-\vec{S}_e(s',\theta')) \cdot \vec{\bar{f}}(s,\theta)  &\approx \int\limits_{-1}^{1} \,d s' \int\limits_{-\pi}^{\pi} \,d \theta' \tensor{K}_{S,e}(s_e(s),s') \cdot \vec{\bar{f}}(s,\theta) \notag \\
 &= 2 \pi \int\limits_{-1}^{1} \,d s'  \tensor{K}_{S,e}(s_e(s),s') \cdot  \vec{\bar{f}}(s,\theta)\,.
 \end{align}
The remaining integral over $s'$ can be evaluated exactly \citep{Gradshteyn2000} to give
\begin{equation}
 2 \pi \int\limits_{-1}^{1} \,d s'  \tensor{K}_{S,e}(s_e(s),s') \cdot  \vec{\bar{f}}(s,\theta) = \tensor{M}_{a} (s,\theta)\cdot  \vec{\bar{f}}(s,\theta)\,, \label{sp_ex}
\end{equation}
where
\begin{align}
\tensor{M}_{a} (s,\theta) &= \left\{\chi_{\parallel}(s_e(s),\theta) \Th(s) \Th(s) +\chi_{\perp}(s_e(s),\theta) \left[\tensor{I}-\Th(s)\Th(s)\right]\right\}\,, \label{Ma}\\
\frac{a}{2\pi}\chi_{\parallel}(s_e,\theta) &=  \frac{1-\beta}{ (-\beta)^{3/2}}L(s_e,\theta)+g(s_e,\theta,1)-g(s_e,\theta,-1)\,, \\
\frac{a}{2\pi} \chi_{\perp}(s_e,\theta) &=  \frac{1}{ \sqrt{-\beta}}L(s_e,\theta)\,, \\
L(s_e,\theta) &= \ln\left(\frac{a(s_e -\beta) + \sqrt{-\beta} \norm{\tilde{\vec{R}}_e(s_e,\theta,-1)}}{a(s_e +\beta) + \sqrt{-\beta} \norm{\tilde{\vec{R}}_e(s_e,\theta,1)}} \right)\,, \\
g(s_e,\theta,s') &= \frac{2 (s_e-s')}{\beta \norm{\tilde{\vec{R}}_e(s_e,\theta,s')}}\left(\frac{  s' s_e \alpha^2 - (1-s_e^2)\beta}{2\beta - s_e^2 (1-\beta) } \right)\,,
\end{align}
and $a$, $\alpha$, $\beta$, and $s_e$ are all also functions of $(s,\theta)$ according to \crefrange{r1}{r6}.

 The last integrand to expand is the mirror singularities that account for the plane interface, $\tensor{G}_{S}^{*}(\vec{S}(s,\theta)-\tensor{A}\cdot\vec{S}(s',\theta'))$. The binomial series approach can also be used to achieve this. Similarly to the free-space Green's function expansion, we express the argument as
\begin{equation}
\vec{S}(s,\theta)-\tensor{A}\cdot\vec{S}(s',\theta') = \vec{R}_0^*(s,s') + \epsilon \Delta\erho^*(s,\theta,s',\theta') , \label{reflect_arg}
\end{equation}
where $\vec{R}_0^*(s,s') = \vec{r}(s) - \tensor{A}\cdot\vec{r}(s) -2 d \Zh$ is a vector between a point on the body centreline and the mirror centreline, $ \Delta\erho^*(s,\theta,s',\theta')  =\rho(s) \erho(s,\theta) - \rho(s') \tensor{A}\cdot\erho(s',\theta')$ is the difference between the cross-section vectors at $(s,\theta)$ and the mirror cross-section vector at $(s',\theta')$, and $\tensor{A}$ is the reflection matrix in $\Zh$. The length squared of the argument can therefore be expressed as
\begin{equation}
    \norm{\vec{S}(s,\theta)-\tensor{A} \cdot \vec{S}(s',\theta')}^2  = \norm{\tilde{\vec{R}}^*(s,s')}^2 \left[1+ R_{\Delta}^{*(2)}(s,\theta,s',\theta')\right]\left[1+ R_{\Delta}^{*(1)}(s,\theta,s',\theta')\right]\,,
\end{equation}
where
\begin{equation}
\norm{\tilde{\vec{R}}^*(s,s')}^2 =  \vec{R}_0^{*2}(s,s') + \epsilon^2 \rho^2(s) + \epsilon^2 \rho^2(s')\,,
\end{equation}
\begin{equation}
\norm{\tilde{\vec{R}}^*(s,s') }^2R_{\Delta}^{*(2)}(s,\theta,s',\theta') = - 2 \epsilon^2 \rho(s) \rho(s') \erho(s,\theta) \cdot \tensor{A}\cdot\erho(s',\theta')\,,
\end{equation}
\begin{equation}
\norm{\tilde{\vec{R}}^*(s,s')}^2 \left[1+ R_{\Delta}^{*(2)}(s,\theta,s',\theta')\right]R_{\Delta}^{*(1)}(s,\theta,s',\theta') = 2\epsilon  \vec{R}_0^*(s,s') \cdot \Delta\erho^*(s,\theta,s',\theta')\,.
\end{equation}
Geometrically, $|\tilde{\vec{R}}^*(s,s')|^2$ is again the total squared lengths of each component vector in \cref{reflect_arg}, while the $R_{\Delta}^{*(i)}(s,\theta,s',\theta')$ contains the interactions between them. Unlike the free space case, $R_{\Delta}^{*(2)}(s,\theta,s',\theta')<1$ for all $(s',\theta')$ because $\norm{\vec{R}_0^*(s,s')} \neq 0$ if the body does not cross the interface. The binomial series, which follow, are therefore always valid. Hence, the denominators in our mirror Green's functions can always be expanded as
\begin{multline}
\norm{\vec{S}(s,\theta)-\tensor{A}\cdot\vec{S}(s',\theta')}^{-2n} = \\\norm{\tilde{\vec{R}}^*(s,s')}^{-2n} \sum_{k_1=0}^{\infty} \binom{-n}{k_1}R_{\Delta}^{*(1)}(s,\theta,s',\theta')^{k_1} \sum_{k_2=0}^{\infty}\binom{-n}{k_2}R_{\Delta}^{*(2)}(s,\theta,s',\theta')^{k_2}\,.
\end{multline}
 Retaining the leading terms, the expansions suggest that we can approximate the mirror singularities with
  \begin{equation}
 \tensor{G}_{S}^*(\vec{S}(s,\theta)-\tensor{A}\cdot\vec{S}(s',\theta')) = \tensor{K}_{S}^*(s,s')+O( R_{\Delta}^{*(i)}(s,\theta,s',\theta'))
 \end{equation}
 for $i=1,2$, where
 \begin{align}
     \tensor{K}_{S}^*(s,s')&=   \left(\frac{\tensor{I}}{\norm{\tilde{\vec{R}}^*}} +\frac{\vec{R}_0^* \vec{R}_0^*}{\norm{\tilde{\vec{R}}^*}^3} \right)\cdot \tensor{B} \notag \\ 
     &{}-  \frac{2 \lambda}{1 + \lambda} (\Zh\cdot\vec{r}(s')-d) (\Zh\cdot\vec{r}(s)-d)\left(\frac{\tensor{I}}{\norm{\tilde{\vec{R}}^*}^3} -3\frac{\vec{R}_0^* \vec{R}_0^*}{\norm{\tilde{\vec{R}}^*}^5}\right)\cdot \tensor{A}  \notag \\
      &{}-  \frac{2 \lambda}{1 + \lambda} (\Zh\cdot\vec{r}(s')-d) \left( \frac{\vec{R}_0^* \Zh - \Zh \vec{R}_0^*}{\norm{\tilde{\vec{R}}^*}^3} \right)\cdot \tensor{A}.
 \end{align}
The leading-order approximation for the integral of the mirror singularities is therefore
\begin{align}
 \int\limits_{-1}^{1} \,d s' \int\limits_{-\pi}^{\pi} \,d \theta' \tensor{G}_{S}^*(\vec{S}(s,\theta)-\tensor{A}\cdot\vec{S}(s',\theta')) \cdot \vec{\bar{f}}(s',\theta')  &\approx \int\limits_{-1}^{1} \,d s' \int\limits_{-\pi}^{\pi} \,d \theta' \tensor{K}_{S}^*(s,s') \cdot \vec{\bar{f}}(s',\theta') \notag \\
 &= 2 \pi \int\limits_{-1}^{1} \,d s'  \tensor{K}_{S}^*(s,s') \cdot \langle \vec{\bar{f}}(s',\theta') \rangle_{\theta'}\,. \label{mirro_ex}
 \end{align}

The expansions of the free space (\cref{sing_ex}), mirror (\cref{mirro_ex}), and regularising spheroid (\cref{sp_ex}) can be combined together to create a leading-order approximation of the regularised boundary integrals, \cref{regG}. This approximation has the form
\begin{equation}
    \mathcal{L}\bar{\vec{f}} = \Delta \tensor{M}_{A}(s,\theta)\cdot \vec{\bar{f}}(s,\theta) +2 \pi \int\limits_{-1}^{1} \,d s' \left( \tensor{K}_{S}(s,s') +\tensor{K}_{S}^*(s,s')\right) \cdot \langle \vec{\bar{f}}(s',\theta') \rangle_{\theta'}\,, \label{lead}
\end{equation}
where $\Delta \tensor{M}_{A}(s,\theta) = \tensor{M}_{A}(s,\theta)- \tensor{M}_{a}(s,\theta)$ is the mobility for the translating spheroid, \cref{BIspheroid}, minus the leading-order representation for this term, \cref{Ma}.  The above integral equation is the leading-order operator that needs to be inverted for the tubular body theory by interfaces expansion. Technically speaking, this integral equation has a compact, symmetric, and self-adjoint kernel which renders the integral equation amenable to analysis and solution. 

Though it involves $(s,\theta)$, the approximate integral equation is actually a one-dimensional Fredholm integral equation of the second kind plus a sequence of linear operations \citep{Koens2022}. The equivalence to a one-dimensional Fredholm integral equation and a sequence of linear operations can be shown by considering the problem $\vec{Q}(s,\theta) = \mathcal{L}\bar{\vec{f}}$. If we multiply this equation by  $\Delta \tensor{M}_{A}^{-1}(s,\theta)$ and then average over $\theta$, it becomes
\begin{multline}
 \langle  \Delta \tensor{M}_{A}^{-1}(s,\theta)\cdot \vec{Q}\rangle_{\theta}   = \langle \vec{\bar{f}}(s,\theta)\rangle_{\theta}\\{}+2 \pi  \langle  \Delta \tensor{M}_{A}^{-1}(s,\theta)\rangle_{\theta}\cdot \int\limits_{-1}^{1} \,d s' \left( \tensor{K}_{S}(s,s') +\tensor{K}_{S}^*(s,s')\right) \cdot \langle \vec{\bar{f}}(s',\theta) \rangle_{\theta} \,, \label{lead2}
\end{multline}
where we have used that $\langle \vec{\bar{f}}(s',\theta') \rangle_{\theta'} =\langle \vec{\bar{f}}(s',\theta) \rangle_{\theta}$.
 The above is a one-dimensional Fredholm integral equation of the second kind for $\langle \vec{\bar{f}}(s,\theta)\rangle_{\theta}$. If this Fredholm integral equation is substituted into $\vec{Q}(s,\theta) = \mathcal{L}\bar{\vec{f}}$, it becomes
 \begin{equation}
    \langle  \Delta \tensor{M}_{A}^{-1}(s,\theta)\rangle_{\theta}\cdot \Delta \tensor{M}_{A}(s,\theta)\cdot \vec{\bar{f}}(s,\theta) = \vec{Q}(s,\theta) - \langle  \Delta \tensor{M}_{A}^{-1}(s,\theta)\cdot \vec{Q}\rangle_{\theta} + \langle \vec{\bar{f}}(s,\theta)\rangle_{\theta}\,. \label{lead3}
 \end{equation}
 This is a linear equation for $\vec{\bar{f}}(s,\theta)$ in terms of $\vec{Q}(s,\theta)$ and $\langle \vec{\bar{f}}(s,\theta)\rangle_{\theta}$. Hence, the leading-order operator of \cref{lead} is equivalent to a one-dimensional Fredholm integral of the second kind with a compact, symmetric, and self-adjoint kernel (\cref{lead2}), plus a sequence of linear operations (\cref{lead3}). Since Fredholm integral equations of the second kind and linear operations are in some sense well behaved, the inversion of the leading-order operator, \cref{lead}, is also expected to behave similarly.
 
 \subsection{Construct the series}

The final step in the tubular-body theory derivation is to represent the  full traction jump in the regularised boundary integrals, $\vec{\bar{f}}(s,\theta)$, as an iterative series, found through repeatedly solving the leading-order operator, \cref{lead}. The simplest approach to achieve this is to add and subtract $\mathcal{L}\bar{\vec{f}}$ from the regularised boundary integrals and rearrange the equation into
\begin{equation}
  8 \pi \vec{U}(\vec{S}(s,\theta)) =  \mathcal{L} \bar{\vec{f}} + \Delta\mathcal{L} \bar{\vec{f}}\,, \label{series_start}
\end{equation}
where $\Delta\mathcal{L} \bar{\vec{f}}$ is the difference between the leading-order operator and right hand side of the regularised boundary integrals and is given by
 \begin{align}
  \Delta\mathcal{L} \bar{\vec{f}} &=  \int\limits_{-1}^{1} \,d s' \int\limits_{-\pi}^{\pi} \,d \theta' \left[ \tensor{G}_{S}(\vec{S}(s,\theta)-\vec{S}(s',\theta'))  - \tensor{K}_{S}(s,s') \right]\cdot \vec{\bar{f}}(s',\theta') \notag \\
 &{}-  \int\limits_{-1}^{1} \,d s' \int\limits_{-\pi}^{\pi} \,d \theta' \left[\tensor{G}_{S}(\vec{S}_{e}(s_e,\theta)-\vec{S}_{e}(s',\theta')) -\tensor{K}_{S,e}(s_e(s),s')\right] \cdot \vec{\bar{f}}(s,\theta) \notag \\
 &{}+\int\limits_{-1}^{1} \,d s' \int\limits_{-\pi}^{\pi} \,d \theta' \left[\tensor{G}_{S}^{*}(\vec{S}(s,\theta)-\tensor{A}\cdot\vec{S}(s',\theta')) - \tensor{K}_{S}^*(s,s') \right]\cdot \vec{\bar{f}}(s',\theta')\,. \label{difference}
\end{align} 
The above expresses the difference terms as integrals to emphasise the relationship between the boundary integral and the leading-order terms.

Since the operator $\mathcal{L}$ should be well behaved, it is reasonable to assume that its inverse exists. Assuming that an inverse $\mathcal{L}^{-1}$ exists,  \cref{series_start} can be written as
\begin{equation}
  8 \pi \mathcal{L}^{-1}\vec{U}(\vec{S}(s,\theta)) =   \left(1+ \mathcal{L}^{-1}\Delta\mathcal{L} \right) \bar{\vec{f}}\,, 
\end{equation}
 where $1 \bar{\vec{f}} = \bar{\vec{f}}$. The solution to the regularised boundary integrals can therefore be written as
\begin{equation}
  \bar{\vec{f}}(s,\theta) =  8 \pi\left(1+ \mathcal{L}^{-1}\Delta\mathcal{L} \right)^{-1} \mathcal{L}^{-1}\vec{U}(\vec{S}(s,\theta)).
\end{equation}
Provided the eigenvalues of $\mathcal{L}^{-1}\Delta\mathcal{L}$ are within $(-1,1)$, which we assume and evidence empirically later, $\left(1+ \mathcal{L}^{-1}\Delta\mathcal{L} \right)^{-1}$ can be expressed as a Neumann series, the operator analogue of a geometric series, allowing the solution to be written as
\begin{equation}
  \bar{\vec{f}}(s,\theta) =  8 \pi \sum_{n=0}^{\infty}\left(-\mathcal{L}^{-1}\Delta\mathcal{L} \right)^{n} \mathcal{L}^{-1}\vec{U}(\vec{S}(s,\theta))
\end{equation}
or, equivalently,
\begin{equation}
  \bar{\vec{f}}(s,\theta) =  \sum_{n=0}^{\infty} \bar{\vec{f}}_n(s,\theta), \label{end1}
\end{equation}
where
\begin{align}
    \mathcal{L} \bar{\vec{f}}_0(s,\theta) &= 8 \pi \vec{U}(\vec{S}(s,\theta))\,, \label{end2} \\
    \mathcal{L} \bar{\vec{f}}_n(s,\theta) &=-\Delta\mathcal{L} \bar{\vec{f}}_{n-1}(s,\theta) \quad n\geq1\,. \label{end3}
\end{align}
\Cref{end1,end2,end3} are the tubular-body theory equations for a body by a plane interface. They are structurally equivalent to TBT for free space \citep{Koens2022}. Identically to the free space version, solutions are constructed by iteratively solving the leading-order kernel. In practice, the iterative approach may be most efficient, depending on the difficulty of inverting the leading-order terms and the number of terms in the series needed to achieve the desired accuracy. We note that, in the previous tubular-body theory study, no condition was identified for the convergence of the series in \cref{end1}. The Neumann series approach used here reveals that the series converges if the eigenvalues of $\mathcal{L}^{-1}\Delta\mathcal{L}$ are within $(-1,1)$, and applies to general operators, not simply those employed in this study.
 
\section{Numerical implementation}
\label{sec: numerical implementation}

The previous TBT study inverted the leading-order operator, $\mathcal{L} \bar{\vec{f}}_n(s,\theta)$ in \cref{end2,end3}, and evaluated the difference integrals, $\Delta\mathcal{L} \bar{\vec{f}}_{n-1}$, through a collocation approach \citep{Koens2022}. This approach was computationally effective as only a few terms in the series, \cref{end1}, was needed. However, it would not be suitable for tubular-body theory by interfaces (TBTi) if significantly more terms are needed, as we will see is often the case. We therefore adopted a Galerkin approach \citep{Pozrikidis1992, Kim2005}, similar to that often applied to the boundary integral equations \citep{Pozrikidis1992}. The Galerkin method to allows us to estimate the eigenvalues of the operator, quickly compute iterations and capture the full solution.

The surface of the tubular body is discretized by dividing $s \in [-1,1]$ and $\theta \in [-\pi,\pi)$ into $N$ and $M$ equal subintervals, respectively. The traction jump is then assumed to be constant over a region of $s \in [s_k-\Delta s/2,s_k+\Delta s/2]$ and $\theta \in [\theta_l- \Delta \theta/2,\theta_l+ \Delta \theta/2)$, where $(s_i,\theta_j)$ is the center of the $(i,j)$th cell on the tubular body and  $\Delta s =2/N$ and $\Delta \theta =2 \pi /M$ are the distance between points in $s$ and $\theta$, respectively. This discretization approximates each surface integral as
\begin{equation}
    \int\limits_{-1}^{1} \,d s' \int\limits_{-\pi}^{\pi} \,d \theta' \vec{Q}(\vec{S}(s_i,\theta_j)-\vec{S}(s',\theta')) \cdot \vec{\bar{f}}(s',\theta') \approx \sum_{k=0}^{N} \sum_{l=0}^{M} \vec{Q}_{i,j,k,l} \cdot \vec{\bar{f}}(s_k,\theta_l)\,, \label{NM}
\end{equation}
where $\vec{Q}(\vec{R})$ represents the integral kernel of \cref{lead} or \cref{difference} and
\begin{equation}
    \vec{Q}_{i,j,k,l} =\int\limits_{s_k-\Delta s/2}^{s_k+\Delta s/2} \,d s' \int\limits_{\theta_l- \Delta \theta/2}^{\theta_l+ \Delta \theta/2} \,d \theta'  \vec{Q}(\vec{S}(s_i,\theta_j)-\vec{S}(s',\theta'))\,. \label{mat}
\end{equation}
 The integrands in \cref{lead,difference} are non-singular, by construction, and are straightforward to evaluate numerically. With the above discretisation, \cref{end2,end3} are transformed into a system of linear equations, which can be represented as the matrix equations
\begin{align}
  \bm{\mathfrak{L}}  \bar{\bm{\mathfrak{f}}}_0 &= 8 \pi \bm{\mathfrak{U}}\,, \label{N1}\\
  \bm{\mathfrak{L}}  \bar{\bm{\mathfrak{f}}}_n &=- \bm{\Delta\mathfrak{L}}  \bar{\bm{\mathfrak{f}}}_{n-1}\,, \quad n\geq1\,, \label{N2}
\end{align}
where $\bm{\mathfrak{U}} = \{\vec{U}(\vec{S}(s_0,\theta_0)),\vec{U}(\vec{S}(s_1,\theta_0)), \dots, \vec{U}(\vec{S}(s_N,\theta_M))\} $ contains the discrete surface velocities and $\bar{\bm{\mathfrak{f}}}_n = \{\bar{\vec{f}}_n(\vec{S}(s_0,\theta_0)),\bar{\vec{f}}_n(\vec{S}(s_1,\theta_0)), \dots, \bar{\vec{f}}_n(\vec{S}(s_N,\theta_M))\}$ is the unknown traction jumps weighted by their respective surface elements. We define the discrete operators $ \bm{\mathfrak{L}}$ and $\bm{\Delta\mathfrak{L}}$ as
\begin{equation}
    \bm{\mathfrak{L}} =
    \begin{pmatrix}
     \mathfrak{L}_{0,0,0,0}    & \mathfrak{L}_{0,0,1,0} & \dots & \mathfrak{L}_{0,0,N,M}  \\
   \mathfrak{L}_{1,0,0,0}    & \mathfrak{L}_{1,0,1,0} & \dots & \mathfrak{L}_{1,0,N,M} \\
   \vdots & \vdots & & \vdots\\ 
    \mathfrak{L}_{N,M,0,0}    & \mathfrak{L}_{N,M,1,0} & \dots & \mathfrak{L}_{N,M,N,M}
    \end{pmatrix}\,,
    \end{equation}
    \begin{equation}
    \bm{\Delta\mathfrak{L}} = \begin{pmatrix}\Delta \mathfrak{L}_{0,0,0,0}    & \Delta \mathfrak{L}_{0,0,1,0} & \dots & \Delta \mathfrak{L}_{0,0,N,M}  \\
  \Delta \mathfrak{L}_{1,0,0,0}    & \Delta \mathfrak{L}_{1,0,1,0} & \dots & \Delta\mathfrak{L}_{1,0,N,M} \\
   \vdots & \vdots & & \vdots\\ 
   \Delta \mathfrak{L}_{N,M,0,0}    & \Delta \mathfrak{L}_{N,M,1,0} & \dots & \Delta \mathfrak{L}_{N,M,N,M}
    \end{pmatrix}\,, 
\end{equation}
as approximations to the full operators $\mathcal{L}$ and $\Delta \mathcal{L}$, with scalar components
\begin{equation}
    \mathfrak{L}_{i,j,k,l} =  \int\limits_{s_k-\Delta s/2}^{s_k+\Delta s/2} \,d s' \int\limits_{\theta_l- \Delta \theta/2}^{\theta_l+ \Delta \theta/2} \,d \theta' \left( \tensor{K}_{S}(s_i,s') +\tensor{K}_{S}^*(s_i,s')\right)  +\Delta \tensor{M}_{A}(s_i,\theta_j) \delta_{i,k} \delta_{j,l}\,,
\end{equation}
\begin{align}
\Delta \mathfrak{L}_{i,j,k,l} &= \int\limits_{s_k-\Delta s/2}^{s_k+\Delta s/2} \,d s' \int\limits_{\theta_l- \Delta \theta/2}^{\theta_l+ \Delta \theta/2} \,d \theta' \left[ \tensor{G}_{S}(\vec{S}(s_i,\theta_j)-\vec{S}(s',\theta'))  - \tensor{K}_{S}(s_i,s') \right] \notag \\
 &{}+\int\limits_{s_k-\Delta s/2}^{s_k+\Delta s/2} \,d s' \int\limits_{\theta_l- \Delta \theta/2}^{\theta_l+ \Delta \theta/2} \,d \theta' \left[\tensor{G}_{S}^{*}(\vec{S}(s_i,\theta_j)-\tensor{A}\cdot\vec{S}(s',\theta')) - \tensor{K}_{S}^*(s_i,s') \right] \notag \\
  &{}-\delta_{i,k} \delta_{j,l} \int\limits_{-1}^{1} \,d s' \int\limits_{-\pi}^{\pi} \,d \theta' \left[\tensor{G}_{S}(\vec{S}_{e}(s_e(s_i),\theta_j)-\vec{S}_{e}(s',\theta')) -\tensor{K}_{S,e}(s_e(s_i),s')\right]\,.
\end{align}
Here,  $\delta_{i,j}$ is the Kronecker delta, defined to be $\delta_{i,j}=1$ when $i=j$ and zero otherwise.

The discretized tubular body theory equations, \cref{N1,N2}, can be solved by inverting $\bm{\mathfrak{L}}$ to find
\begin{align}
   \bar{\bm{\mathfrak{f}}}_0 &= 8 \pi \bm{\mathfrak{L}} ^{-1}  \bm{\mathfrak{U}}\,, \\
  \bar{\bm{\mathfrak{f}}}_n &=\left(- \bm{\mathfrak{L}} ^{-1}  \bm{\Delta\mathfrak{L}} \right)\bar{\bm{\mathfrak{f}}}_{n-1}\,, \quad n\geq1\,.
\end{align}
It is useful to retain the full  $- \bm{\mathfrak{L}} ^{-1} \bm{\Delta\mathfrak{L}}$ matrix as it reduces the task of finding higher iterations of $\bar{\bm{\mathfrak{f}}}_n $ to matrix multiplication. 

This matrix representation also allows us to compute the infinite summation using the aforementioned Neumann series for matrices, which generalises the well-known geometric series to operators. Specifically,
\begin{equation}
  \bar{\bm{\mathfrak{f}}} =  \sum_{n=0}^{\infty}  \bar{\bm{\mathfrak{f}}}_n = 8 \pi \sum_{n=0}^{\infty} \left(- \bm{\mathfrak{L}} ^{-1}  \bm{\Delta\mathfrak{L}} \right)^n \bm{\mathfrak{L}} ^{-1}  \bm{\mathfrak{U}} = 8 \pi (1+\bm{\mathfrak{L}} ^{-1}  \bm{\Delta\mathfrak{L}} )^{-1} \bm{\mathfrak{L}} ^{-1}  \bm{\mathfrak{U}}, \label{full}
\end{equation}
if the eigenvalues of $\bm{\mathfrak{L}} ^{-1}  \bm{\Delta\mathfrak{L}}$ lie in $(-1,1)$. These eigenvalues are an approximation to the eigenvalues of ${\mathcal{L}} ^{-1}  {\Delta\mathcal{L}}$. Hence, the Galerkin approach can be used to determine the tubular-body theory by interfaces solution, estimate the eigenvalues of  ${\mathcal{L}} ^{-1}  {\Delta\mathcal{L}}$, and test the convergence of the series representation for the traction jump, \cref{end1}, with relative ease.
 
 \section{Evaluation for a spheroid by a plane wall} \label{sec:comparison}

We numerically compared the results tubular-body theory by interfaces to boundary element computations for the case of a spheroid by a plane wall. The plane wall configuration was chosen because rigid body motion close to a plane wall generates large stresses from lubrication effects. The stresses from lubrication diverge as the separation between the wall and the body decreases, making them difficult to resolve numerically \citep{Kim2005}. The symmetry axis of the spheroid was taken to be perpendicular to the wall normal in order to coincide with previously published data \citep{Koens2021a}. In this configuration, the spheroid can be parameterised as
\begin{equation}
\vec{S}(s,\theta) = s \vec{\hat{x}} + \epsilon \sqrt{1-s^2} \erho -d \Zh, 
\end{equation}
where we have set $\vec{r}(s) = s \vec{\hat{x}}$ and $\rho(s) = \sqrt{1-s^2}$. 

 \subsection{Rigid body motion from boundary element simulations} \label{sec:boundary element}

\begin{table}
    \centering
    \begin{tabular}{|c|c c c|} \hline
         & $\epsilon=1$& $\epsilon=0.2$ & $\epsilon=0.1$ \\\hline
      $d=2$   & $ \begin{bmatrix}
       R^{FU}_{11} \to 1.4\% (0.34)     \\
       R^{FU}_{22} \to   1.3\% (0.33) \\ 
        R^{FU}_{33} \to  1.2\% (0.48)
      \end{bmatrix}$ &  $ \begin{bmatrix}
       1.3\% (0.10)     \\
          1.5\% (0.15) \\ 
        1.5\% (0.18)
      \end{bmatrix}$ & $\begin{bmatrix}
       1.5\% (0.08)     \\
          1.6\% (0.12) \\ 
        1.6\% (0.14)
      \end{bmatrix}$   \\
      $d=2 \epsilon$ &  $ \begin{bmatrix}
       1.4\% (0.34)     \\
        1.3\% (0.33) \\ 
        1.2\% (0.48)
      \end{bmatrix}$ &  $ \begin{bmatrix}
       1.3\% (0.14)     \\
          1.4\% (0.23) \\ 
        1.2\% (0.44)
      \end{bmatrix}$ & $ \begin{bmatrix}
       1.3\% (0.12)     \\
          1.4\% (0.22) \\ 
        1.1\% (0.44)
      \end{bmatrix}$\\
      $d=1.1 \epsilon$ & $ \begin{bmatrix}
       1.2\% (0.50)     \\
          1.1\% (0.47) \\ 
        0.2\% (0.46)
      \end{bmatrix}$ & $ \begin{bmatrix}
       0.9\% (0.19)     \\
         0.9\% (0.31) \\ 
        0.4\% (1.55)
      \end{bmatrix}$ & $ \begin{bmatrix}
       1.0\% (0.18)     \\
         0.9\% (0.32) \\ 
        1.3\% (4.62)
      \end{bmatrix}$\\ \hline
    \end{tabular}
    \caption{The relative error in the non-zero force-translation resistance coefficients $\{R^{FU}_{11},R^{FU}_{22},R^{FU}_{33}\}$. The absolute errors between the coefficients is given in parentheses.}
    \label{tab:A}
\end{table}

\begin{table}
    \centering
    \begin{tabular}{|c|c c c|} \hline
         & $\epsilon=1$& $\epsilon=0.2$ & $\epsilon=0.1$ \\\hline
      $d=2$   & $ \begin{bmatrix}
       R^{L\Omega}_{11} \to 2.2\% (0.56)     \\
       R^{L\Omega}_{22} \to   2.2\% (0.56) \\ 
        R^{L\Omega}_{33} \to  2.2\% (0.55)
      \end{bmatrix}$ &  $ \begin{bmatrix}
       2.3\% (0.02)     \\
          2.4\% (0.11) \\ 
        2.4\% (0.11)
      \end{bmatrix}$ & $ \begin{bmatrix}
       2.4\% (0.004)     \\
          2.5\% (0.08) \\ 
        2.5\% (0.08)
      \end{bmatrix}$   \\
      $d=2 \epsilon$ &  $ \begin{bmatrix}
       2.2\% (0.56)   \\
        2.2\% (0.56) \\ 
        2.2\% (0.55)
      \end{bmatrix}$ &  $ \begin{bmatrix}
       2.3\% (0.02)     \\
          2.1\% (0.19) \\ 
        2.3\% (0.13)
      \end{bmatrix}$ & $ \begin{bmatrix}
       2.3\% (0.004)     \\
          1.9\% (0.17) \\ 
        2.3\% (0.11)
      \end{bmatrix}$\\
      $d=1.1 \epsilon$ & $ \begin{bmatrix}
       2.1\% (0.77)     \\
          2.1\% (0.76) \\ 
        2.2\% (0.60)
      \end{bmatrix}$ & $ \begin{bmatrix}
       2.1\% (0.027)     \\
         0.1\% (0.027) \\ 
        1.9\% (0.16)
      \end{bmatrix}$ & $ \begin{bmatrix}
       2.2\% (0.007)     \\
         0.9\% (0.33) \\ 
        1.8\% (0.14)
      \end{bmatrix}$\\ \hline
    \end{tabular}
    \caption{The relative error in the non-zero torque-rotation resistance coefficients $\{R^{L\Omega}_{11},R^{L\Omega}_{22},R^{L\Omega}_{33}\}$. The absolute error between the coefficients is given in parentheses.}
    \label{tab:C}
\end{table}

\begin{table}
    \centering
    \begin{tabular}{|c|c c c|} \hline
         & $\epsilon=1$& $\epsilon=0.2$ & $\epsilon=0.1$ \\\hline
      $d=2$   & $ \begin{bmatrix}
       R^{F\Omega}_{12} \to 6.0\% (0.0073)     \\
       R^{F\Omega}_{21} \to   3.2\% (0.0040) 
      \end{bmatrix}$ &  $ \begin{bmatrix}
       1.7\% (0.0017)     \\
         2.5\% (10^{-6}) 
      \end{bmatrix}$ & $ \begin{bmatrix}
       1.7\% (9\times10^{-4})      \\
        6.8\% (2 \times 10^{-7})
      \end{bmatrix}$   \\
      $d=2 \epsilon$ &  $ \begin{bmatrix}
        6.0\% (0.0073)     \\
        3.2\% (0.0040) 
      \end{bmatrix}$ &  $ \begin{bmatrix}
       0.7\% (0.0056)     \\
          16.5\% (9 \times 10^{-4})
      \end{bmatrix}$ & $ \begin{bmatrix}
       0.5\% (3 \times 10^{-4})     \\
          91.6\% (5 \times 10^{-4}) 
      \end{bmatrix}$\\
      $d=1.1 \epsilon$ & $ \begin{bmatrix}
       5.8\% (0.12)     \\
          4.2\% (0.09) 
      \end{bmatrix}$ & $ \begin{bmatrix}
       6.6\% (0.20)     \\
         64.3\% (0.02) 
      \end{bmatrix}$ & $ \begin{bmatrix}
       15.1\% (0.2765)     \\
         402.9\% (0.0082) 
      \end{bmatrix}$\\ \hline
    \end{tabular}
    \caption{The relative error in the non-zero force-rotation resistance coefficients $\{R^{F\Omega}_{12},R^{F\Omega}_{21}\}$. The absolute error between the coefficients is given in parentheses.}
    \label{tab:B}
\end{table}

The eight non-zero elements of the resistance matrix found by our implementation of TBTi through \cref{full} were compared to the coefficients found using full boundary element simulations, considering inverse aspect ratios of $\epsilon=1,0.2,0.1$ and distances of $d = 2, 2 \epsilon, 1.1 \epsilon$. The TBTi calculations used $\lambda=10^4$, $N=15$ and $M=300$, whilst the boundary element calculations discretised the body into \SI{2e4} flat triangles and numerically solved \cref{eq: boundary integral equations} as described by \citet{Walker2018} and \citet{Pozrikidis2002a}. These particular configurations were chosen so as to include cases when the particle is far from the wall ($d=2$), close to the wall ($d=2\epsilon$), and very close to the wall ($d=1.1 \epsilon$). The inverse aspect ratios were chosen to range from spheres ($\epsilon=1$), to increasingly prolate spheroids ($\epsilon=0.2,0.1$). In each case, the three non-zero force-translation (\cref{tab:A}), three non-zero torque-rotation (\cref{tab:C}), and two non-zero force-rotation coefficients (\cref{tab:B}) were determined and the absolute error and relative error computed. The absolute error is defined to be the absolute difference between the two coefficients and the relative error is defined as the absolute error divided by the boundary element simulation result.

Irrespective of the inverse aspect ratio and the distance from the wall, the non-zero force-translation (\cref{tab:A}) and torque-rotation (\cref{tab:C}) resistance coefficients display a small relative error between the TBTi simulations and the boundary element method, with these errors attributed to the differing discretisations. The lack of dependence on distance or the inverse aspect ratio suggests that the TBTi approach handles shapes well beyond the slender limit and captures their lubrication behaviour correctly. 

The force-rotation resistance coefficients (\cref{tab:B}) display a larger and apparently more erratic error behaviour. This is because of the size of these terms are much smaller, making them more sensitive to numerical error. The fact that the relative errors tend to be larger when the coefficients are smaller supports this. The error in the numerical integration, used to construct $\bm{\mathfrak{L}}$ and $ \bm{\Delta\mathfrak{L}}$ in the TBTi program, was set to $10^{-6}$ throughout and may be responsible for the error observed, while the boundary element method used for verification did not have configurable error tolerances.

 \subsection{Rigid body translations from regularised boundary element simulations} 

\begin{figure}
    \centering
   \includegraphics[width=0.95\textwidth]{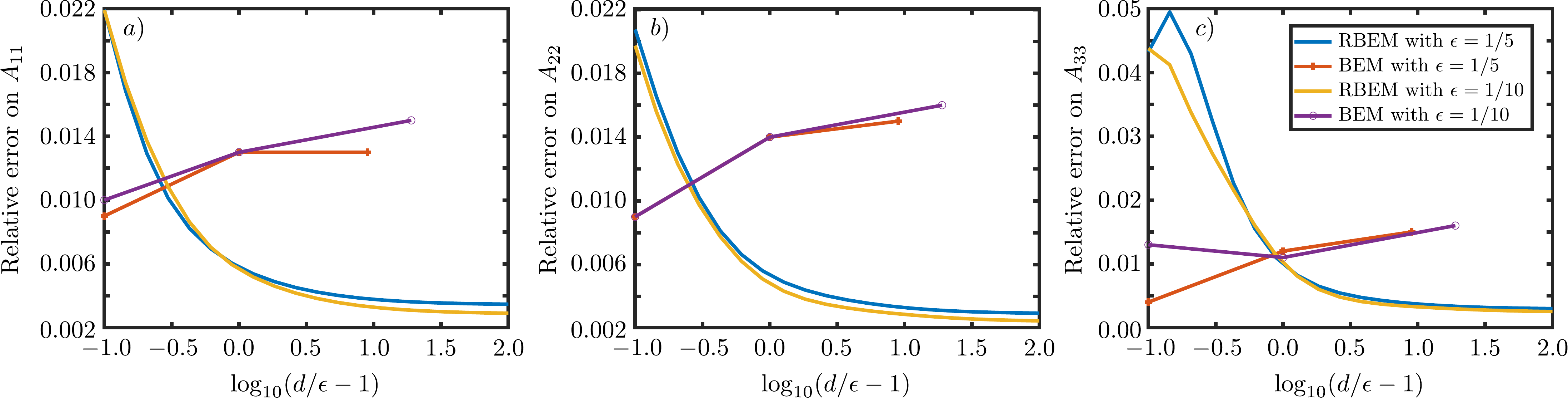}
    \caption{The relative error between between TBTi and regularised boundary element method (RBEM) simulations for the translation of a spheroid as a function of the distance to the wall. The relative error between TBTi and the normal boundary element method (BEM) are included for reference. }
    \label{fig: RBEM}
\end{figure}

Regularised boundary element simulations for a prolate spheroid translating near a plane wall were developed to test the accuracy of a local drag theory for rods above a plane wall \citep{Koens2021a}. The regularisation parameter for these simulations was set to $\epsilon/5$ for each simulation. These regularised boundary element simulations were also compared to the TBTi representation (Fig.~\ref{fig: RBEM}). The TBTi simulations used $\lambda=10^4$, $N=10$, and $M=100$ for this comparison.

Tubular-body theory by interfaces demonstrates less than 1\% relative error with the regularised boundary integral simulations when the spheroid is far from the wall. However, the error appears to increase as the gap size, $\Delta d= d-\epsilon$, decreases below $\epsilon$ to around 2\% error for motions perpendicular to the wall normal and 5\% for motions parallel to the wall normal for gap spaces of $0.1\epsilon$. The relatively large error for near-wall motion is surprising, as normal boundary element method simulations predicted a 1-2\% relative error with the TBTi for the same distances. The discrepancy is because the regularised boundary element method and the normal boundary element method provide different predictions when close to the wall. Since the normal boundary element method is an approximation to the exact boundary integrals and the implementation used was created specifically to model filaments close to walls \citep{Walker2018}, the difference suggests that the regularised method, with its regularisation of $\epsilon/5$, was not able to maintain accuracy when within $\epsilon$ of the wall, which is not entirely unexpected. 

 \subsection{Eigenvalues}

\begin{figure}
    \centering
   \includegraphics[width=0.9\textwidth]{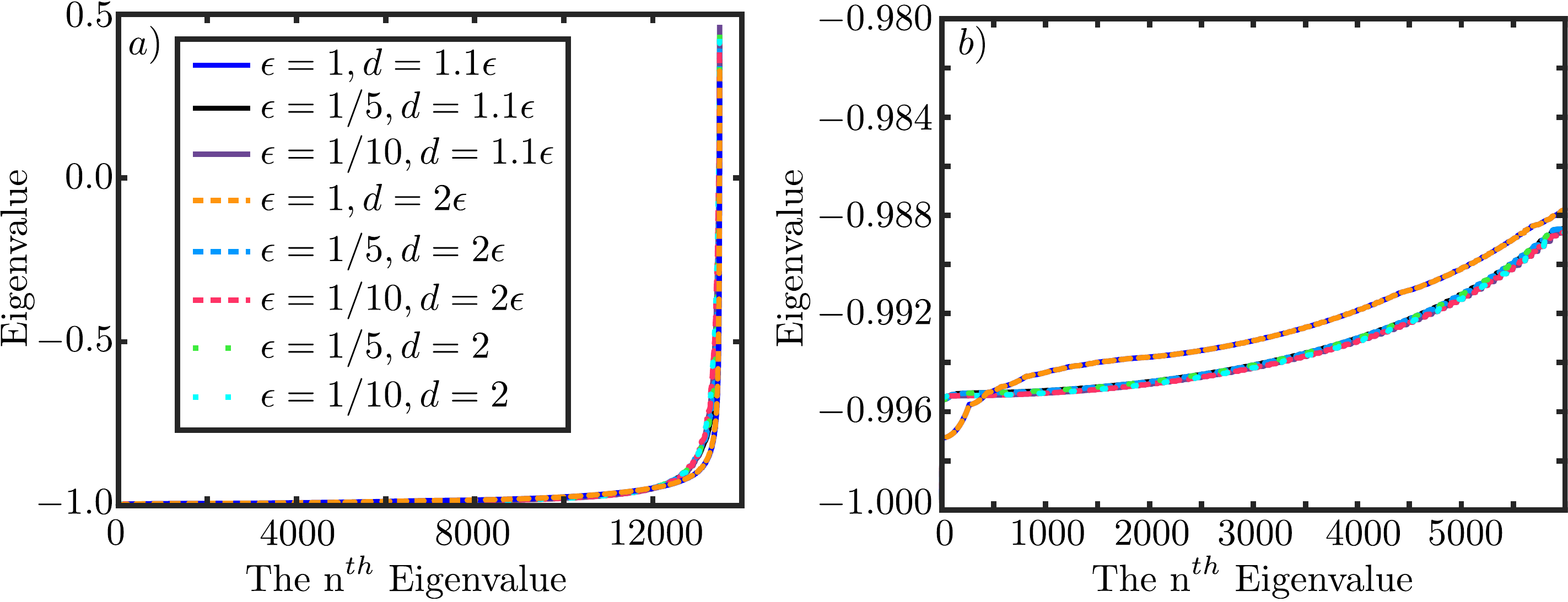}
    \caption{The eigenvalues of the matrix approximation to the tubular body theory by interfaces operator, $\bm{\mathfrak{L}} ^{-1} \cdot \bm{\Delta\mathfrak{L}}$, for the eight distinct configurations considered in \cref{sec:boundary element}. a) shows all 135,000 eigenvalues. b) shows the 6,000 smallest eigenvalues. }
    \label{fig: eig}
\end{figure}

 The series representation used in TBTi, \cref{end1}, requires the eigenvalues of $\mathcal{L}^{-1} \Delta \mathcal{L}$ to lie between $(1,-1)$ to converge absolutely. The eigenvalues of this operator can be approximated via the eigenvalues of the matrix approximation to this operator, $\bm{\mathfrak{L}} ^{-1} \cdot \bm{\Delta\mathfrak{L}}$. Hence, the nature of the eigenvalues of $\bm{\mathfrak{L}} ^{-1} \cdot \bm{\Delta\mathfrak{L}}$ were explored for the eight different configurations in Sec.~\ref{sec:boundary element} (Fig.~\ref{fig: eig}). 
 
 In all configurations, the eigenvalues of $\bm{\mathfrak{L}} ^{-1} \cdot \bm{\Delta\mathfrak{L}}$ decrease rapidly from around 0.5 before plateauing to value just above -1. The inverse aspect ratio, $\epsilon$, slightly slows the rate at which this transition occurs. However, the distance to the wall, $d$, had surprisingly little effect. The average difference between eigenvalues for the same $\epsilon$ and different $d$ is $0.015$ or smaller. The behaviour seen suggests that the eigenvalues of $\mathcal{L}^{-1} \Delta \mathcal{L}$ also lie within $(-1,1)$, and that the TBTi series, \cref{end1}, converges absolutely. 

 \subsection{Convergence rates}

\begin{figure}
    \centering
   \includegraphics[width=0.95\textwidth]{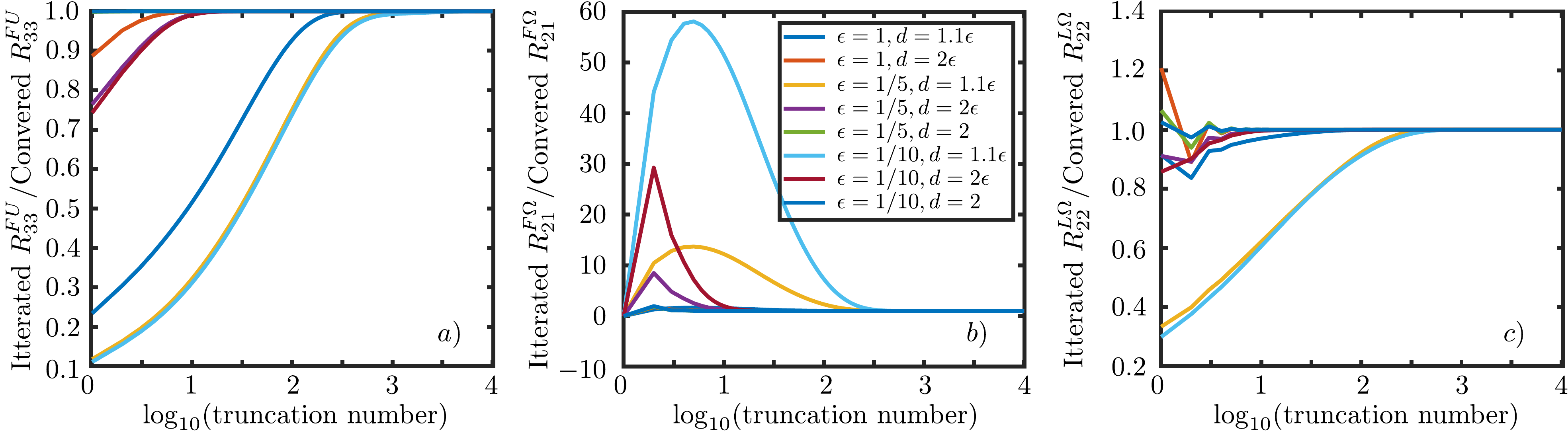}
    \caption{The convergence of the non-zero resistance coefficients predicted by the TBTi series, \cref{end1}, as a function of the truncation point in the series. The slowest coefficients to converge, in each sub-matrix, is shown for brevity. a) The coefficient relating force towards the wall from motion towards the wall, $R^{FU}_{33}$. b) The coefficient relating force along the minor axis perpendicular to the wall normal from rotation around the major axis, $R^{F\Omega}_{21}$. c) The coefficient relating torque in the minor axis perpendicular to the wall normal from rotation in the same direction, $R^{L\Omega}_{22}$. All coefficients are scaled by their converged value. }
    \label{fig: converge}
\end{figure}

The convergence of the tubular-body theory by interfaces summation, \cref{full}, as a function of the number of terms retained, was also explored for the eight different configurations in \cref{sec:boundary element} (\cref{fig: converge}). For brevity, only the force towards the wall from motion in the same direction, $R^{FU}_{33}$, force in the minor axis perpendicular to the wall due to rotation around the major axis, $R^{F\Omega}_{21}$, and the torque in the minor axis perpendicular to the wall normal from rotation in the same direction, $R^{L\Omega}_{22}$, are shown because they converge the slowest. The force-translation and torque-rotation both correspond to motions with large lubricating stresses, while the force-rotation term is a small secondary effect of the wall.

Except when the body is very close to the wall, the coefficients are seen to converge in approximately 10 terms. A similar number of terms was needed to realise convergence in the free space TBT \citep{Koens2022}. The number of terms needed to converge increases rapidly when the body is very close to the wall ($d=1.1 \epsilon$), and slightly as the inverse aspect ratio, $\epsilon$, decreases. For the weaker lubrication singularity, present in $R^{L\Omega}_{22}$ (\cref{fig: converge} c), convergence occurs around 100 terms, while for the strongest lubrication singularity, present in $R^{FU}_{33}$ (\cref{fig: converge} a), it takes approximately 1,000 terms to converge. The small coupling term, $R^{F\Omega}_{21}$, converges in roughly $10^{2.5} \approx 316$ terms. The singular nature of lubrication effects often makes these singularities hard to resolve numerically, so the increase in the number of terms needed for convergence is expected.
 
 \section{Traction jump on helix above an interface} \label{sec:helix}

Tubular-body theory by interfaces works for general cable-like bodies by any plane interface. For example, it can be used to determine the traction jump on a helix moving close to a free interface and a plane wall. We parameterise a helix by $\rho(s) = \sqrt{1 - s^{20}}$ and $\vec{r} = r_x\vec{\hat{x}} + r_y\vec{\hat{y}} + r_z\Zh$, where
\begin{subequations}
\begin{align}
    r_x(s) &=\alpha_h s\,,\\
    r_y(s) &= R_h\cos(ks + \pi/2)\,,\\
    r_y(s) &= R_h\sin(ks + \pi/2)\,,
\end{align}
\end{subequations}
 $\alpha_h = \Lambda / \sqrt{\pi^2 R_{h}^2 + \Lambda^2}$ is the axial length of the helix $k =\pi / \sqrt{\pi^2 R_{h}^2 + \Lambda^2}$ is the wave-number, $R_{h}$ is the helix radius and $\Lambda$ is the helix pitch.  The helix-by-an-interface simulations here used $\epsilon = 0.05$, $R_h = 0.05109375$, and $\lambda= 0.25$. This parametrisation was used to simulate the motion of tightly wound helices with the free-space TBT \citep{Koens2022}. The specific geometry corresponds to the helix with the largest pitch and smallest helix radius tested by \citet{Koens2022}. When the distance from the interface was large, $d=1,000$, the results found using TBTi and the free space TBT were the same, up to numerical error.

\begin{figure}
    \centering
    \begin{overpic}[width=0.9\textwidth]{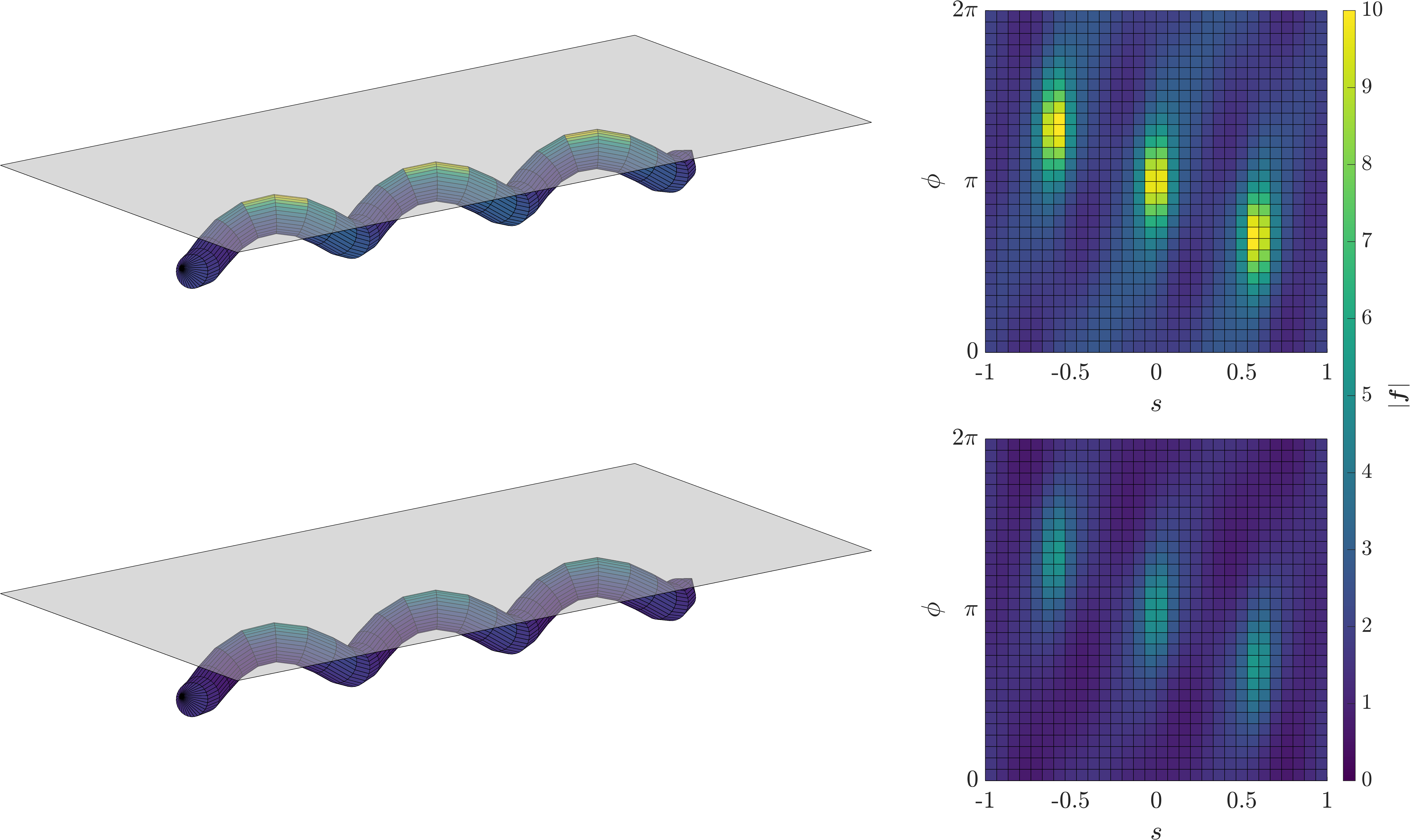}
        \put(0,60){(a)}
        \put(60,60){(b)}
        \put(0,30){(c)}
        \put(60,30){(d)}
        \linethickness{1pt}
        \put(31,49){\vector(0,1){7}}
    \end{overpic}
    \caption{The traction jump on a helical body as it approaches an infinite plane boundary. The colour shows the magnitude of the traction computed using tubular-body theory by interfaces for two identical helical bodies moving towards a rigid boundary (a,b) with $\lambda \rightarrow \infty$ and a free interface (c,d) with $\lambda = 0$. In (b) and (d), we show the same traction distributions in the computational domain, from which we observe significant differences between different parts of each body and between the two cases. The approach towards the rigid boundary is associated with significantly larger traction, as expected.}
    \label{fig: helix}
\end{figure}
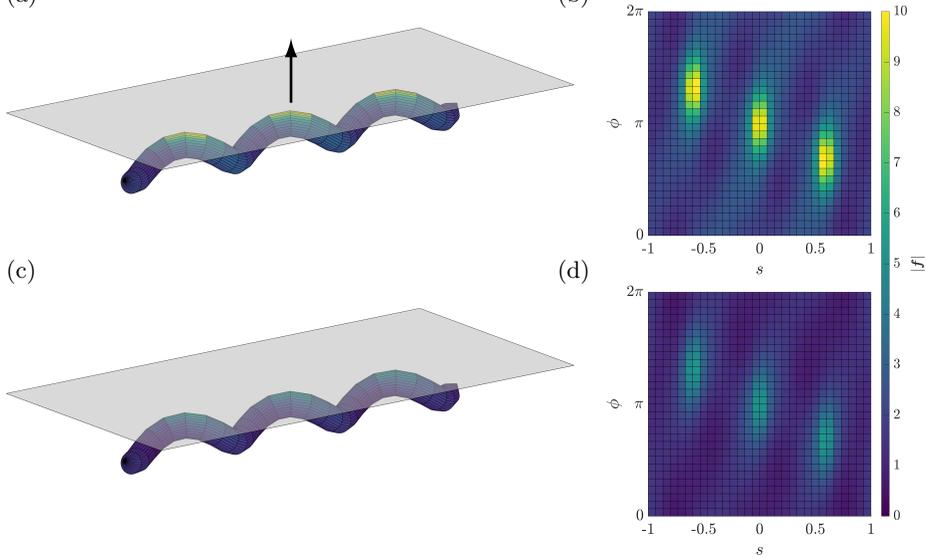

The surface traction on this helix was determined in presence of a rigid boundary ($\lambda\rightarrow\infty$) and a free interface ($\lambda = 0$), using the TBTi formalism. The distance to the wall was $d = 0.15$. Each configuration is illustrated in \cref{fig: helix}, with \cref{fig: helix}a and \cref{fig: helix}c corresponding to the rigid boundary and the free interface held flat by surface tension, respectively. In both cases, we prescribe a unit velocity towards the boundary on the surface of the helix, and colour the surface by the pointwise magnitude of the resulting traction jump (multiplied by the surface element), with the boundaries shown semi-transparent for visual clarity. 

 The traction distribution on the computational domain, parameterised by arclength $s$ and angle $\phi$, is shown in \cref{fig: helix}b and \cref{fig: helix}d. The largest-magnitude traction jumps (multiplied by the surface element) are found on the three near-boundary regions of the helix in both cases. The decay of these peaks are skewed along the helix arms, giving the curving shape on the computational domain.
 
 The rigid boundary is seen to generate tractions jumps (multiplied by the surface element) about twice as large than the free interface in these regions. Since the traction jump multiplied by the surface element scales with the total force on the body, this is consistent with the known behaviour of the lubrication force. When approaching another body, the lubrication force diverges proportionally with the inverse of the gap size, $\Delta d$ \citep{Kim2005}. The force on the nearest points of the helix by the wall therefore scales with $1/\Delta d$. However, a helix approaching a free interface is mathematically equivalent to the helix approaching a mirrored helix across the interface. Hence, the effective gap size for the helix approaching the plane interface is doubled. The traction jump on a helix by a free interface therefore scales with $1/(2 \Delta d)$. This difference explains the apparent factor of 2 observed in the traction strengths and implies that TBTi can handle complex shapes by different types of interfaces.

\section{Conclusion} \label{sec: conclusion}

This paper extends the tubular-body theory formalism to handle cable-like bodies by plane interfaces. Similarly to in the free-space case, the employed expansion allows for the traction jump on the body to be reconstructed exactly by iteratively solving a better-behaved slender-body theory-like operator, \cref{lead}. The iterations are shown to be equivalent to an appropriate analogue of the geometric series, indicating that the iterations will converge to the exact value if certain conditions on the eigenvalues of the operator are met. Empirically, these conditions were found to be satisfied for all geometries considered.

The tubular-body theory by interfaces equations (\cref{end1,end2,end3}) were solved numerically using a Galerkin approach \citep{Pozrikidis1992}, resulting in an efficient method to conduct iterations, determine the exact solution, and find approximate eigenvalues for the system. The TBTi simulations were compared to normal and regularised boundary element simulations for spheroids by a plane wall. All rigid body motions near a wall generate lubrication stresses that can be hard to determine numerically. The TBTi results agreed well with both boundary element simulations for all aspect ratios and distances from the wall, thereby suggesting that the method captures the lubrication effect. The largest deviations between the results were found in the weak force-rotation resistance coefficients and was likely due to the numerical errors in both the TBTi and boundary element method implementations. 

The TBTi equations were found to converge in around 10 iterations when the body was well separated from the boundary. However, when very close to the wall, the rate of convergence decreased. When a body approaches the plane wall it was found to converge in around 1,000 terms, while for other motions it took around 100 terms. The increase in the number of terms reflects the general difficulty with resolving lubrication effects numerically.

Finally, the TBTi simulations were used to look at the motion of a helix towards a rigid wall and a free interface. As would be anticipated, the traction (multiplied by the surface element) found in both cases was largest on the parts of the helix closest to the interface and decayed as the distance increased. The maximum traction on the helix near a plane wall was also found to be around twice the size of the maximum traction on the helix by a free interface. Since the hydrodynamics of a body by a free interface is equivalent to two bodies approaching each other at double the separation, the factor of two is consistent with the scaling of the lubrication singularity.

The TBTi formalism opens up many new possibilities for exploration. It allows a slender-body theory-like method to explore  geometries that lie well beyond the limits of slender-body theory. Looking forward, the convergence rate of the representation could be improved if a better regularizing body could be found. The derivation could also generalise to other systems that can be represented by integral equations, and other viscous flow configurations. Furthermore, the well-behaved nature of the TBTi operator opens up new avenues for solving for the hydrodynamics of wires near interfaces asymptotically.

 Declaration of Interests: The authors report no conflict of interest.

BJW is supported by the Royal Commission for the Exhibition of 1851. 

The TBTi program and data that support the findings of this study are openly available on GitHub at \url{https://github.com/LKoens/TBTi}. 

\bibliographystyle{jfm}
\bibliography{library}
\end{document}